\documentclass[reprint,amsmath,amssymb,aps,pra,onecolumn]{revtex4-2}
\usepackage[utf8]{inputenc}
\usepackage{amsmath}
\usepackage{pgfplots}
\pgfplotsset{compat=newest}
\usepackage{xcolor}
\usepackage{tikz}
\usepackage{tkz-euclide}
\usepackage{comment}
\usepackage[T1]{fontenc}
\usetikzlibrary{decorations.pathmorphing,3d}
\begin{document}

\title{Weakly interacting Bose gases in the canonical ensemble}
\author{Jonata S. Soares$^{1,2}$}\thanks{jownaplido34@usp.br}
\author{Axel Pelster$^{2}$}\thanks{axel.pelster@rptu.de}
\author{Arnaldo Gammal$^{1}$}\thanks{gammal@if.usp.br}

\affiliation{$^{1}$Instituto de F\'{i}sica, Universidade de S\~{a}o Paulo, 05508-090 S\~{a}o Paulo, Brazil.\\ 
$^{2}$Department of Physics and Research Center OPTIMAS, RPTU University Kaiserslautern-Landau, Erwin-Schrödinger Straße 46, 67663 Kaiserslautern, Germany.}

\begin{abstract}
Based on the canonical description of a non-interacting Bose gas, we
work out how both thermodynamic and statistical properties change
perturbatively with respect to weak two-particle interactions. Up to
first order, we obtain a recursion formula for the canonical partition
function, which consists of the same Feynman diagrams as the grand-canonical description but with different Feynman rules. Resumming
this recursion formula for the canonical partition function allows one
to characterize the statistics of the ground-state occupancy by its respective cumulants. We demonstrate the applicability of this approach
by analyzing a dilute Bose gas with contact interaction in a box trap. To this end, we used Dirichlet
boundary conditions in view of their relevance for current experiments
with atomic gases, where the box trap is implemented, for instance,
with digital mirror devices. 
\end{abstract}

\date{\today}
\maketitle

\section{Introduction}

The first experimentally realized Bose-Einstein condensates of dilute alkali gases were systems with particle numbers ranging from thousand to hundred thousand particles \cite{ketterle1995, cornell}. Since then, many efforts have been made to study atomic Bose gases with an even larger number of particles in order to improve the signal-to-noise ratio \cite{PhysRevLett.81.3811, large-atom-number}. This experimental development justifies that many theoretical calculations for Bose-Einstein condensates are performed in the thermodynamic limit with an infinitely large number of particles in an infinitely large spatial confinement. Then the finiteness of the real experiment is taken into account by determining finite-size corrections of that thermodynamic limit \cite{kluender}. 

Theoretically, the corresponding calculations can be performed with different ensembles \cite{huang, rzazewski2025}. In order to statistically describe a certain experiment, one should obviously choose that ensemble whose assumptions correspond best to the situation in the laboratory. Quite often in theory the grand-canonical ensemble is used as there the thermodynamic limit is most straight-forwardly accessible and usually the predictions of all ensembles agree in the thermodynamic limit. However, there are important exceptions to that thumb rule. For instance, at absolute zero the particle number fluctuations of the ground state vanish in the canonical ensemble and are reduced for an increasing particle number. However, the grand-canonical ensemble yields at absolute zero non-vanishing particle number fluctuations of the ground state, which even increase linearly with the particle number \cite{ziff,kirsten,scully}. One may be tempted to consider such huge fluctuations in the grand-canonical ensemble as unphysical. However, seminal experiments with photon gases in dye-filled microcavities \cite{photon-nature} confirmed those striking predictions of the grand-canonical ensemble \cite{photon-BEC}. Due to many photon absorptions and emissions by the dye molecules,  this system  realizes not only an energy but also a particle bath for the photons in the microcavity. In contrast to this, more recent experiments with atomic gases indicate that they are better described by a canonical or even a microcanonical ensemble \cite{arlt2021,rzazewski2023}. 

Another line of research deals with the intriguing question how small particle numbers could be in order to still detect indications of macrosopic quantum phenomena. For instance, a
driven-dissipative non-equilibrium Bose-Einstein condensation was realized for less than 10 photons \cite{nyman}. Furthermore, the emergence of Cooper pairs was observed in a mesoscopic two-dimensional Fermi gas, which consisted of just 12 atoms \cite{preiss}. And the quantum critical behaviour at the many-body localization transition was detected in a disordered Bose–Hubbard system of 12 lattice sites at unity filling \cite{greiner}. From all these studies one can read off that it is difficult to determine when a physical system has reached sufficient size for its macroscopic properties to be well described by many-body theory. However, there is experimental evidence that increasing the number of particles one by one yields a fast convergence of observables toward the many-body limit \cite{preiss}.

In view of ongoing dilute gas experiments with a quite small number of atoms, there is a need to develop theoretical tools for their canonical description. Historically, the canonical treatment focused mainly on dealing with the indistinguishability of particles at the expense of neglecting any two-particle interaction. For an ideal quantum gas, all particles are represented by cycles winding around a cylinder in Euclidean space-time whose circumference is the imaginary-time $\hbar\beta$ with the reciprocal temperature $\beta = 1/(k_{\rm B}T)$. Taking into account the cycle decomposition of the permutation group then leads to the seminal recursion formula for the partition function with respect to the number of particles \cite{landsberg,feynman,borrmann,sato,wilkens,brosens,ivanov,kleinert,glaum2007}. Iterating this recursion formula can be done quite efficiently up to one million particles.
However, including two-particle interactions in canonical calculations is much more difficult due to a significant increase of combinatorial complexity. Some initial attempts revealed that a plain perturbative approach for the canonical partition function is problematic for small enough temperatures \cite{tempere2000,glaummaxplanck}.
Therefore,
although the canonical description of non-interacting bosonic gases is well developed in the literature, only few articles deal with the corresponding case of weak two-particle interactions \cite{navez1998,scully2000,xiong}. This motivates us in this article to systematically extend the previous findings for non-interacting bosons in the canonical ensemble in view of including weak interactions perturbatively in the first order. 
To this end, we invoke the perturbation theory, which is applicable for weak interactions, i.e.~a small gas parameter. Thus, we neglect the impact of quantum fluctuations, but we deal with thermal fluctuations. Note that the impact of quantum fluctuations on condensate statistics is studied through Bogoliubov theory. Some studies apply this approach even in the canonical ensemble, see e.g.~Refs.~\cite{wang,Idziaszek_2009, scully2000}, which deal with a Bose gas in a finite box with periodic boundary condition. 

In this paper, we analyze systematically weakly interacting Bose gases in the canonical ensemble. We start in Sect.~II with reviewing the cycle decomposition of the permutation group for ideal Bose gases in the canonical ensemble, where we deal first with the $N$-particle partition function and then with the $N$-particle density matrix. The latter turns out through its normalization to provide a recursive formula for the partition function. We extend this study in Sect.~III for a weakly interacting Bose gas using  perturbation theory up to first order for a canonical ensemble. With this we obtain a diagrammatic representation of the partition function with the same Feynman diagrams as in the grand-canonical ensemble but with different Feynman rules for the canonical ensemble.
However, this yields a perturbative recursion formula that suffers from a negative partition function for low enough temperatures. Therefore, we work out an appropriate resummed recursion formula, which we then apply in Sect.~IV. With this, we obtain for a three-dimensional Bose gas in a box trap with Dirichlet boundary conditions and contact interaction a thermodynamic and statistical description in the canonical ensemble.
\section{Non-interacting canonical ensemble}
We start with describing a non-interacting Bose gas in
the canonical ensemble. To this end, we work out how to calculate the partition function, the density matrix, and the probability of a ground-state occupancy.
\subsection{Partition function}
The partition function $Z_{N}^{(0)}(\beta)$ for a gas of $N$ non-interacting bosons as a function of the inverse temperature $\beta=1/ k_{\rm B} T$
has the path integral representation \cite{kleinert}
\begin{eqnarray}
Z_{N}^{(0)}(\beta) = \frac{1}{N!} \sum_{\text{P}} \int d\mathbf{x}_{1} \cdots \int d\mathbf{x}_{N}  \prod_{n=1}^{N} \int_{\mathbf{x}_{n}(0) = \mathbf{x}_{n}}^{\mathbf{x}_{n}(\hbar \beta) = \mathbf{x}_{\text{P}(n)}} \mathcal{D}\mathbf{x}_{n}(\tau)\;e^{-\mathcal{A}^{(0)}[\mathbf{x}_1, \ldots, \mathbf{x}_N]/\hbar} \,,   
    \label{can1}
\end{eqnarray}
where the permutation P takes into account the indistinguishability of the particles and
\begin{equation}
 \mathcal{A}^{(0)}[\mathbf{x}_1, \ldots, \mathbf{x}_N] = \sum_{j=1}^{N} \int_{0}^{\hbar \beta} \hspace*{-2mm} d\tau \left\{ \frac{M}{2} \dot{\mathbf{x}}_{j}^{2}(\tau) + V(\mathbf{x}_j(\tau))\right\}    
  \label{can2}
\end{equation}
represents the corresponding action. Note that it is assumed here that all bosons are exposed to the same external potential $V(\mathbf{x})$.
Evaluating the path integral (\ref{can1}) on the imaginary-time paths $\mathbf{x}_{1}(\tau), 
  \ldots , \mathbf{x}_{N}(\tau)$ yields \cite{kleinert}
\begin{eqnarray}
Z_{N}^{(0)}(\beta) = \frac{1}{N!} \sum_{\text{P}} \int d\mathbf{x}_{1} \cdots \int d\mathbf{x}_{N}  \left( \mathbf{x}_{\text{P}(1)}, \ldots , \mathbf{x}_{\text{P}(N)},\hbar \beta | 
   \mathbf{x}_{1}, \ldots , \mathbf{x}_{N} ,0 \right)^{(0)}\, .
    \label{can3}
\end{eqnarray}
Here the integrand denotes the $N$-particle imaginary-time propagator, which factorizes due to the assumed absence of any interaction according to
\begin{equation}
  \hspace*{-2mm}  \left( \mathbf{x}_{\text{P}(1)}, ... , \mathbf{x}_{\text{P}(N)},\hbar \beta 
    |\mathbf{x}_{1}, ... , \mathbf{x}_{N},0 \right)^{(0)} = \prod_{j=1}^{N} \left(  \mathbf{x}_{\text{P}(j)}, \hbar \beta 
    |\mathbf{x}_{j} ,0\right)^{(0)} \,.
    \label{can5}
\end{equation}
Each one-particle imaginary-time propagator has the spectral decomposition
\begin{equation}
 \left(  \mathbf{x}, \tau    |\mathbf{x}' ,\tau'\right)^{(0)} =
\sum_{\mathbf{n}} \psi_{\mathbf{n}}^*(\mathbf{x})
e^{- \varepsilon_{\mathbf{n}}(\tau - \tau')/\hbar} \psi_{\mathbf{n}} (\mathbf{x}')
\label{can5c}
\end{equation}
with $\psi_{\mathbf{n}}(\mathbf{x})$ and $\varepsilon_{\mathbf{n}}$ denoting 
orthonormal energy eigenfunctions and eigenvalues with the quantum number $\mathbf{n}$. These spectral data follow from the eigenvalue problem 
\begin{equation}
\hat{H}^{(0)} \psi_{\mathbf{n}} (\mathbf{x}) = \varepsilon_{\mathbf{n}} \psi_{\mathbf{n}} (\mathbf{x})
\label{can5d}
\end{equation}
of the one-particle Hamilton operator
\begin{equation}
\hat{H}^{(0)} = - \frac{\hbar^2}{2m}\,\Delta + V({\mathbf x})\, .
\label{can5e}
\end{equation}
Combining Eqs.~(\ref{can3}) and (\ref{can5}) yields
\begin{eqnarray}
    Z_{N}^{(0)}(\beta) = \frac{1}{N!} \sum_{\text{P}} \int d\mathbf{x}_{1} \cdots \int d\mathbf{x}_{N} \,  (\mathbf{x}_{\text{P}(1)}, \hbar \beta | \mathbf{x}_{1},0)^{(0)} \cdots (\mathbf{x}_{\text{P}(N)}, \hbar \beta | \mathbf{x}_{N},0)^{(0)} ~.
    \label{can6}
\end{eqnarray}
Evaluating this $N$-particle partition function for an ideal Bose gas requires a proper treatment of the permutation sum. To this end, we use the fact that the symmetric permutation group $S_{N}$ for bosons relies on a cycle decomposition \cite{feynman}. Each permutation P is decomposed into $C_n$
 cycles of length $n$, so that the constraint $N= \sum_n nC_n$ holds. Then the contribution of one cycle of length $n$ to the partition function (\ref{can6}) reads
\begin{eqnarray}
h_{n}(\beta) = \int d\mathbf{x}_1 \cdots \int d\mathbf{x}_n\;
    (\mathbf{x}_{1}, \hbar \beta | \mathbf{x}_{n},0)^{(0)}  (\mathbf{x}_{n}, \hbar \beta | \mathbf{x}_{n-1},0)^{(0)}\times \cdots \times (\mathbf{x}_{2}, \hbar \beta | \mathbf{x}_{1},0)^{(0)}\, .
    \label{can8}
\end{eqnarray}
This is evaluated by taking into account both the translational invariance with respect to the imaginary time
\begin{eqnarray}
(\mathbf{x}_{1}, \tau_1 | \mathbf{x}_2,\tau_2)^{(0)}   =  (\mathbf{x}_{1}, \tau_1- \tau_2 | \mathbf{x}_2,0 )^{(0)}
    \label{can8a}
\end{eqnarray}
and the group property of the one-particle imaginary-time propagator
\begin{eqnarray}
(\mathbf{x}_{1}, \tau_1 | \mathbf{x}_2,\tau_2)^{(0)}   = \int d\mathbf{x} \,   (\mathbf{x}_{1}, \tau_1 | \mathbf{x},\tau)^{(0)}  (\mathbf{x}, \tau | \mathbf{x}_2,\tau_2)^{(0)}\, ,
    \label{cangroup}
\end{eqnarray}
following from the spectral decomposition
(\ref{can5c}) and the orthonormality of the energy eigenfunctions. The result reads
\begin{eqnarray}
h_{n}(\beta) = Z_1(n\beta)\, ,
    \label{can8c}
\end{eqnarray}
where the one-particle partition function is defined via
\begin{eqnarray}
Z_1(\beta) = \sum_{\mathbf{n}} e^{- \beta \varepsilon_{\mathbf{n}}}\, .
\label{can8b}
\end{eqnarray}
Thus, the canonical partition function (\ref{can6}) reduces to
\begin{eqnarray}
     && Z_{N}^{(0)}(\beta) = \frac{1}{N!} \sum_{\text{P}}
\prod_{n=1}^{\sum n C_n=N} \Big[ Z_1(n\beta) \Big]^{C_n}\, .
    \label{can6b}
\end{eqnarray}
In this decomposition, there are multi-particle interferences due to these cycles, as discussed in \cite{buchleitner}. However, instead of considering $C_n$ as a function of the permutation P, we can also inversely 
represent the permutation P as a combination of different cycles $C_n$ of length $n$.
Thus, the summation over all permuations P in (\ref{can6b}) then proceeds into 
a summation over all $N$-tuples  $(C_1,\ldots , C_N)$ fulfilling the constraint 
$\sum_n n C_n=N$. 
However, it must be taken into account that a fixed $N$-tuple  $(C_1,\ldots , C_N)$ could appear multiple times. The corresponding combinatorics finally yield the cycle decomposition 
\begin{eqnarray}
  Z_{N}^{(0)}(\beta) = \sum_{C_1, ..., C_N}^{\sum nC_n = N} \prod_{n=1}^{N} \frac{1}{C_n!} \left[ \frac{Z_{1}(n \beta)}{n}\right]^{C_n}\, .
    \label{can9}
\end{eqnarray}
To evaluate (\ref{can9}) for a concrete system is computationally demanding for larger particle numbers $N$ as it requires determining all cycle numbers $C_n$ obeying the constraint $\sum_n n C_n=N$. 
In the following, we will show that the $N$-particle partition function $Z_{N}^{(0)}(\beta)$ also follows alternatively from a recursive procedure, which turns out to be more efficient for concrete calculations.
\subsection{Density matrix}\label{dens-zero}
Whereas the partition function allows one to determine all global thermodynamic properties of interest, local observables like correlation functions 
follow from the density matrix. To this end, we define now the canonical 
one-particle density matrix with two coordinates $\mathbf{x}_1$ and $\mathbf{x}'_1$ for $N$ particles following Ref.~\cite{glaumthesis}. The path integral
representation
\begin{eqnarray}
\hspace*{-7mm}\rho_{N}^{(0)}(\mathbf{x}_1, \mathbf{x}'_1 ; \beta) = \frac{1}{N! Z_{N}^{(0)}(\beta)} \sum_{\text{P}} \int d\mathbf{x}_{2} \cdots \int d\mathbf{x}_{N}  \int_{\mathbf{x}_{1}(0) = \mathbf{x}_{1}'}^{\mathbf{x}_{1}(\hbar \beta) = \mathbf{x}_{\text{P}(1)}} \hspace*{-1mm}\mathcal{D}\mathbf{x}_{1}(\tau)
\prod_{n=2}^{N} 
\int_{\mathbf{x}_{n}(0) = \mathbf{x}_{n}}^{\mathbf{x}_{n}(\hbar \beta) = \mathbf{x}_{\text{P}(n)}} \hspace*{-1mm}\mathcal{D}\mathbf{x}_{n}(\tau)\;e^{-\mathcal{A}^{(0)}[\mathbf{x}_1, \ldots, \mathbf{x}_N]/\hbar}  
    \label{can10c}
\end{eqnarray}
immediately yields
\begin{eqnarray}
\rho_{N}^{(0)}(\mathbf{x}_1, \mathbf{x}'_1 ; \beta) = \frac{1}{N! Z_{N}^{(0)}(\beta)} \sum_{\text{P}} \int d\mathbf{x}_2 \cdots \int d\mathbf{x}_N  \left( \mathbf{x}_{P(1)}, \mathbf{x}_{\text{P}(2)},\ldots , \mathbf{x}_{\text{P}(N)},\hbar \beta | 
   \mathbf{x}_{1}', \mathbf{x}_2, \ldots , \mathbf{x}_{N} ,0 \right)^{(0)}\, .\label{can10} 
\end{eqnarray}

Thus, the difference in defining $\rho^{(0)}_{N}(\mathbf{x}_1, \mathbf{x}'_1 ; \beta)$ in (\ref{can10}) and $Z_{N}^{(0)}(\beta)$ in (\ref{can3})
is that we exclude one coordinate from integration for the one-particle density matrix and fix its initial and final positions 
${\mathbf x}_1$ and $\mathbf{x}_1'$, respectively. Therefore, we conclude that the diagonal one-particle density matrix (\ref{can10}) is normalized by definition due to (\ref{can3}):
\begin{equation}
    \int d\mathbf{x}_{1}\;\rho_{N}^{(0)}(\mathbf{x}_1, \mathbf{x}_1 ; \beta) = 1\, .
    \label{can15}
\end{equation}
Since we assumed to consider no interaction between the particles, taking into account (\ref{can5}) converts (\ref{can10}) to
\begin{eqnarray}
 \hspace{-5mm}\rho_{N}^{(0)}(\mathbf{x}_1, \mathbf{x}'_1 ; \beta) = \frac{1}{N! Z_{N}^{(0)}(\beta)} \sum_{\text{P}} \int d\mathbf{x}_2 \cdots \int d\mathbf{x}_N   (\mathbf{x}_{P(1)}, \hbar \beta |\mathbf{x}_1', 0)^{(0)}  (\mathbf{x}_{\text{P}(2)}, \hbar \beta |\mathbf{x}_2, 0)^{(0)} \cdots (\mathbf{x}_{\text{P}(N)}, \hbar \beta |\mathbf{x}_N, 0)^{(0)}.
 \label{can10b}
\end{eqnarray}
Also here a cycle of length $k$ may appear, which contributes via 
\begin{eqnarray}
    g_{k}(\mathbf{x}_1, \mathbf{x}_1' ; \beta) = \int d\mathbf{x}_2 \cdots \int d\mathbf{x}_k \; (\mathbf{x}_1, \hbar \beta|\mathbf{x}_k, 0)^{(0)}  (\mathbf{x}_k, \hbar \beta|\mathbf{x}_{k-1}, 0)^{(0)} \cdots \,(\mathbf{x}_3, \hbar \beta|\mathbf{x}_2, 0)^{(0)} (\mathbf{x}_2, \hbar \beta|\mathbf{x}'_1, 0)^{(0)}\, . 
    \label{can11} 
\end{eqnarray}
Note that (\ref{can11}) is similar to (\ref{can8}) but while in $h_{n}(\beta)$ the initial and final positions are the same, so a closed cycle occurs between $\mathbf{x}_1$ and $\mathbf{x}_n$,
we have that $g_{k}(\mathbf{x}_1, \mathbf{x}_1' ; \beta)$ an open cycle with intermediate positions between $\mathbf{x}_2$ and $\mathbf{x}_k$ as well as fixed end points $\mathbf{x}_1$ and $\mathbf{x}_1'$. Using again
the translational invariance and the group property of the one-particle imaginary-time propagator (\ref{can5c}), Eq.~(\ref{can11}) reduces to
\begin{equation}
    g_{k}(\mathbf{x}_1, \mathbf{x}_1' ; \beta) = (\mathbf{x}_1, k \hbar \beta | \mathbf{x}'_1, 0)^{(0)}\, .
    \label{can12}
\end{equation}
Thus, the one-particle density matrix (\ref{can10b}) decomposes analogously 
to the partition function (\ref{can6b})
into contributions of cycles of different lengths. But now we have to
take into account not only closed cycles with (\ref{can8c}) but also open cycles with (\ref{can12}), whereas in the partition function all cycles are closed. With a corresponding combinatorial consideration, we arrive at
\cite{glaumthesis}
\begin{eqnarray}
\rho_{N}^{(0)}(\mathbf{x}_1, \mathbf{x}'_1 ; \beta) = \frac{1}{N\,Z_{N}^{(0)}(\beta)}\sum_{k=1}^N
(\mathbf{x}_1, k \hbar \beta | \mathbf{x}'_1, 0)^{(0)} \sum_{C_1, ..., C_N}^{\sum nC_n = N-k} \prod_{n=1}^{N-k} \frac{1}{C_n!} \left[ \frac{Z_{1}(n \beta)}{n}\right]^{C_n}\, .
\label{can13}
\end{eqnarray}
Here we recognize that the cycle decomposition of a partition function (\ref{can9}) for $N-k$ particles appears, so we finally end up with the following result for the one-particle density matrix:
\begin{eqnarray}
    \rho_{N}^{(0)}(\mathbf{x}_1, \mathbf{x}'_1 ; \beta) = \frac{1}{N\,Z_{N}^{(0)}(\beta)} 
    \sum_{k=1}^{N} \;(\mathbf{x}_1, k \hbar \beta | \mathbf{x}'_1, 0)^{(0)}\,Z_{N-k}^{(0)}(\beta) ~.
    \label{can14}
\end{eqnarray}
Due to the normalization property (\ref{can15}) for the one-particle density matrix (\ref{can14}) and the spectral representation of the one-particle imaginary-time propagator (\ref{can5c}), we conclude
that the $N$-particle partition function obeys the recursion relation
\begin{equation}
     Z_{N}^{(0)}(\beta) = \frac{1}{N} \sum_{k=1}^{N} Z_{1}(k \beta)\,Z_{N-k}^{(0)}(\beta) \, .
    \label{can16}
\end{equation}
With Eq.~(\ref{can16}) the $N$-particle partition function can be determined recursively from smaller to larger particle numbers. To this end, only one needs
the one-particle partition function $Z_{1}(\beta)$ defined by its spectral decomposition (\ref{can8b}) and the starting value of the partition function in the absence of any particles is given by 
$Z_{0}^{(0)}(\beta) = 1$. 
\subsection{Ground-state occupancy}\label{2C}
There is another important statistical quantity, which can be extracted from the one-particle density matrix in the canonical ensemble. Namely one can ask the question how to determine the probability $p_{\mathbf{q}}^{(0)}(k|N,\beta)$ of finding $k$ particles in a given state $\mathbf{q}$ at the inverse temperature $\beta$ for $N$ particles. Specializing the state $\mathbf{q}$ to the ground state, this then yields the ground-state occupancy, which represents for bosons the analog of the condensate fraction in the canonical ensemble.

To answer this question, we revisit the recursion relation for the $N$-particle partition function $Z_N^{(0)}(\beta)$ in (\ref{can16}). It contains the one-particle
partition function (\ref{can8b}), which can be decomposed according to
\begin{eqnarray}
Z_1(\beta)=\gamma_{\mathbf{q}}(\beta)+\xi_{\mathbf{q}} (\beta)
\label{can19}
\end{eqnarray}
in the contribution 
\begin{eqnarray}
\gamma_{\mathbf{q}}(\beta)=e^{- \beta \varepsilon_{\mathbf{q}}} 
\label{can20}
\end{eqnarray}
of a specific state $\mathbf{q}$ and the rest
\begin{eqnarray}
\xi_{\mathbf q} (\beta)= \sum_{{\mathbf{n}} \neq \mathbf{q}} e^{- \beta \varepsilon_{\mathbf {n}}}\, . 
\label{can21}
\end{eqnarray}
Inserting (\ref{can19}) into (\ref{can16}) yields
\begin{equation}
     Z_{N}^{(0)}(\beta) = \frac{1}{N} \sum_{k=1}^{N} \Big[\gamma_{\mathbf{q}}(k\beta)+\xi_{\mathbf{q}} (k\beta)
    \Big] \,Z_{N-k}^{(0)}(\beta) \,,
    \label{can22}
\end{equation}
which allows us to unravel the statistics of the specific state $\mathbf{q}$. Namely, since (\ref{can20}) obeys the property $\gamma_{\mathbf{q}} (k \beta) = \gamma_{\mathbf{q}}^{k}(\beta)$, we read off
that
\begin{equation}
    W^{(0)}_{\mathbf{q}}(k|N,\beta) = \gamma_{\mathbf{q}}^{k}(\beta)\,Z_{N-k}^{(0)}(\beta)
    \label{can23}
\end{equation}
represents the weight of finding {\it at least} $k$ particles in state $\mathbf{q}$. Thus, the weight of having {\it exactly} $k$ particles in state $\mathbf{q}$ is given by
\begin{equation}
    P^{(0)}_{\mathbf{q}}(k|N,\beta) = W^{(0)}_{\mathbf{q}}(k|N,\beta) - W^{(0)}_{\mathbf{q}}(k+1|N,\beta)\, ,
    \label{can24}
\end{equation}
so the corresponding probability reads
\begin{equation}
    p_{\mathbf{q}}^{(0)}(k|N,\beta) = \frac{P_{\mathbf{q}}^{(0)}(k|N,\beta)}{Z_{N}^{(0)}(\beta)}\, .
    \label{can25}
\end{equation}
Finally, combining (\ref{can23})--(\ref{can25}), we determine that the probability of finding $k$ particles in a specific state $\mathbf{q}$  follows from \cite{wilkens, glaumthesis}:
\begin{equation}
    p_{\mathbf{q}}^{(0)}(k|N,\beta) = \gamma_{\mathbf{q}}^{k}(\beta)\,\frac{Z_{N-k}^{(0)}(\beta)}{Z_{N}^{(0)}(\beta)} - \gamma_{\mathbf{q}}^{k+1}(\beta)\,\frac{Z_{N-k-1}^{(0)}(\beta)}{Z_{N}^{(0)}(\beta)} \,.
    \label{can26}
\end{equation}
Based on this probability any moment follows according to
\begin{equation}
\langle k^l \rangle_{\mathbf{q}}^{N,\beta}= \sum_{k=1}^N k^l\,
    p_{\mathbf{q}}^{(0)}(k|N,\beta)  \,.
    \label{can27a}
\end{equation}
The zeroth moment turns out to correspond to normalization $\langle 1 \rangle_{\mathbf q}^{N,\beta}=1$, the first moment yields
\begin{equation}
\langle k \rangle_{\mathbf{q}}^{N,\beta}=  \sum_{k=1}^N  \gamma_{\mathbf{q}}^{k}(\beta)\,\frac{Z_{N-k}^{(0)}(\beta)}{Z_{N}^{(0)}(\beta)}\, ,
    \label{can28a}
\end{equation}
and the second moment is given by
\begin{equation}
\langle k^2 \rangle_{\mathbf{q}}^{N,\beta}=  \sum_{k=1}^N  (2k-1) \gamma_{\mathbf{q}}^{k}(\beta)\,
\frac{Z_{N-k}^{(0)}(\beta)}{Z_{N}^{(0)}(\beta)}\, .
\label{can29a}
\end{equation}
Identifying the specific state ${\mathbf q}$ with the ground state ${\mathbf n}_{\rm G}$, the average number of bosons in the ground state corresponds to $\langle k \rangle_{{\mathbf n}_{\rm G}}^{N,\beta}$, while the variance of the ground-state fluctuations follows from $\langle k^2 \rangle_{\mathbf{n}_{\rm G}}^{N,\beta} - \left[\langle k \rangle_{\mathbf {n}_{\rm G}}^{N,\beta}\right]^2$.
\section{Weakly interacting canonical ensemble theory}
After having dealt with the non-interacting Bose gas in the canonical ensemble, we work here the impact of a weak two-particle interaction. Thus, we work out here within first-order perturbation theory the partition function, the density matrix, and the probability for a ground-state occupancy.
\subsection{Partition function}\label{part-weak}
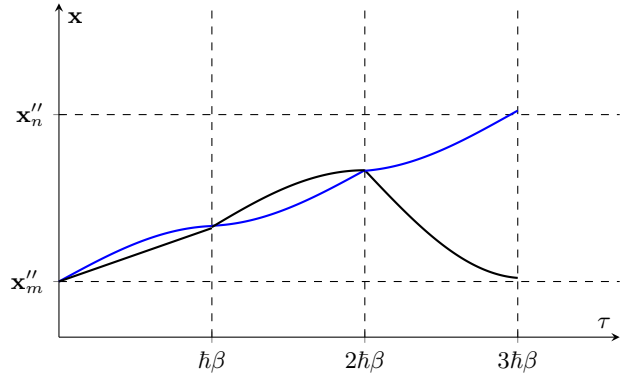
\begin{figure}

\begin{tikzpicture}
\begin{axis}[
    axis lines=middle,
    xlabel={$\tau$},
    ylabel={$\textbf{x}$},
    xtick={0, 1.5, 3.0, 4.5},
    xticklabels={$0$, $\hbar\beta$, $2\hbar\beta$, $3\hbar\beta$},
    ytick={0.5, 2},
    yticklabels={$\textbf{x}_{m}''$, $\textbf{x}_{n}''$},
    xmin=0, xmax=5.5,
    ymin=0, ymax=3,
    samples=200,
    domain=0:10,
    width=9cm,
    height=6cm,
    clip=false
]

\draw[dashed] (axis cs:1.5,0) -- (axis cs:1.5,3);
\draw[dashed] (axis cs:3.0,0) -- (axis cs:3.0,3);
\draw[dashed] (axis cs:4.5,0) -- (axis cs:4.5,3);
\draw[dashed] (axis cs:0.0,0.5) -- (axis cs:5.5,0.5);
\draw[dashed] (axis cs:0.0,2.0) -- (axis cs:5.5,2.0);

\addplot[blue, thick, domain=0:1.5] {0.5*sin(deg(x)) + 0.5};   
\addplot[black, thick, domain=0:1.5] {0.32*x+0.5};
\addplot[blue, thick, domain=1.5:3.0] {0.5*sin(deg(x - 3)) + 1.5};
\addplot[black, thick, domain=1.5:3.0] {0.55*cos(deg(x - 3)) + 0.95} ;
\addplot[black, thick, domain=3.0:4.5] {-0.97*sin(deg(x - 3)) + 1.5} ; 
\addplot[blue, thick, domain=3.0:4.5] {0.51*cos(deg(x - 6)) + 2.0} ;

\end{axis}
\end{tikzpicture}

\caption{Example of two interacting cycles of length $k=3$: Direct path (black) of Eq.~(\ref{H3}) has the same initial and final interaction point, whereas exchange path (blue) of Eq.~(\ref{H3b}) has different initial and final interaction points.}
\label{Fig1}
\end{figure}

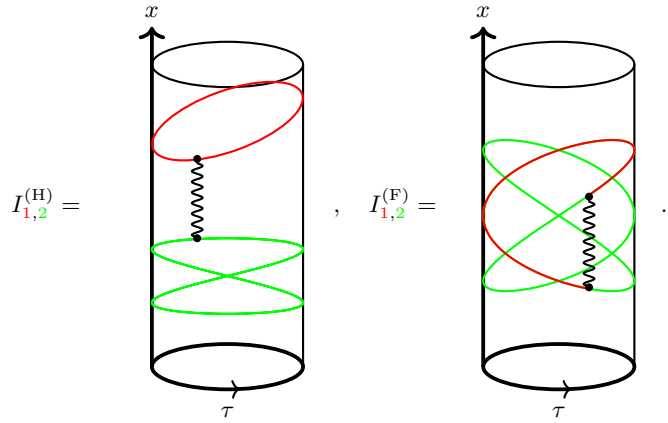
\begin{figure}
\[
I^{({\rm H})}_{\textcolor{red}{1}, \textcolor{green}{2}} =
\quad \quad
\begin{tikzpicture}[baseline={(current bounding box.center)}] 

\draw[thick] (0,0) ellipse (1cm and 0.3cm);
\draw[thick] (0,4) ellipse (1cm and 0.3cm);

\draw[thick] (-1,0) -- (-1,4);
\draw[thick] (1,0) -- (1,4);

\draw[->,line width=0.5mm ] (-1,0) -- (-1, 4+0.5) node[above] {$x$};
\draw[-latex, line width=0.5mm ] (0,0.0) ellipse (1cm and 0.3cm);
\draw[thick, ->, domain=-200:280, samples=100, smooth, variable=\t]
    plot ({cos(\t)*1}, {0.0 + sin(\t)*0.3});
\node at (0.,-0.6) {$\tau$};
\draw[thick, green, domain=0:720, samples=200, smooth, variable=\t]
plot ({sin(2*\t)}, {1.2+0.5*sin(1*\t)});
\draw[thick, red, rotate around={20:(0,3.25)}] (0,3.25) ellipse (1.05cm and 0.4cm);

\draw[thick, decorate, decoration={snake, amplitude=2pt, segment length=4pt}]
      (-0.4,1.70) -- (-0.4,2.75);

\fill (-0.4,1.70) circle (1.5pt);
\fill (-0.4,2.75) circle (1.5pt);

\end{tikzpicture}
\quad , \quad
I^{({\rm F})}_{\textcolor{red}{1},\textcolor{green}{2}} = 
\quad
\begin{tikzpicture}[baseline={(current bounding box.center)}]

\draw[thick] (0,0) ellipse (1cm and 0.3cm);
\draw[thick] (0,4) ellipse (1cm and 0.3cm);

\draw[thick] (-1,0) -- (-1,4);
\draw[thick] (1,0) -- (1,4);

\draw[->,line width=0.5mm ] (-1,0) -- (-1, 4+0.5) node[above] {$x$};
\draw[-latex, line width=0.5mm ] (0,0.0) ellipse (1cm and 0.3cm);
\draw[thick, ->, domain=-200:280, samples=100, smooth, variable=\t]
    plot ({cos(\t)*1}, {0.0 + sin(\t)*0.3});
\node at (0.,-0.6) {$\tau$};
\draw[thick, green, domain=126:486, samples=200, smooth, variable=\t]
  plot ({sin(3*\t)}, {2+sin(2*\t)});

\draw[thick, red, domain=8:128, samples=400, smooth, variable=\t]
  plot ({sin(3*\t)}, {2+sin(2*\t)});

\draw[thick, decorate, decoration={snake, amplitude=2pt, segment length=4pt}]
      (0.4,1.05) -- (0.4,2.25);

\fill (0.4,1.05) circle (1.5pt);
\fill (0.4,2.25) circle (1.5pt);

\end{tikzpicture}
\quad .
\]
\caption{Feynman cylinder representation of the Hartree and Fock energies for three particles corresponding to Eqs.~(\ref{can45}) and (\ref{can46}). The Hartree energy corresponds to two closed cycles of lengths $1$ and $2$, between which an interaction occurs, whereas the Fock energy represents one closed cycle of length $3$, where an open cycle of length $1$ interacts within another open cycle of length $2$.}
\label{Fig2}
\end{figure}

The partition function for an interacting system of $N$ bosons has a similar representation as the non-interacting one (\ref{can1}):
\begin{eqnarray}
Z_{N}(\beta) = \frac{1}{N!} \sum_{\text{P}} \int d\mathbf{x}_{1} \cdots \int d\mathbf{x}_{N}  \prod_{n=1}^{N}  \int_{\mathbf{x}_{n}(0) = \mathbf{x}_{n}}^{\mathbf{x}_{n}(\hbar \beta) = \mathbf{x}_{\text{P}(n)}} \mathcal{D}\mathbf{x}_{n}(\tau)\;e^{-\mathcal{A}[\mathbf{x}_1, \ldots, \mathbf{x}_N]/\hbar} \, .
    \label{can27}
\end{eqnarray}
Here the action 
\begin{equation}
\hspace*{-5mm}    \mathcal{A}[\mathbf{x}_1, \ldots, \mathbf{x}_N] = \mathcal{A}^{(0)}[\mathbf{x}_1, \ldots, \mathbf{x}_N] + \mathcal{A}^{(\text{int})}[\mathbf{x}_1, \ldots, \mathbf{x}_N] 
    \label{can28}
\end{equation}
decomposes into the non-interacting action (\ref{can2}) and the interacting contribution
\begin{equation}
    \mathcal{A}^{(\text{int})}[\mathbf{x}_1, ..., \mathbf{x}_N] = \frac{1}{2} \sum_{n \neq m = 1}^{N} \int_{0}^{\hbar \beta} d\tau\;V^{(\text{int})}\Big(\mathbf{x}_{n}(\tau) - \mathbf{x}_{m}(\tau)\Big) \, ,
    \label{can29}
\end{equation}
where $V^{({\rm int})}({\mathbf x}-{\mathbf x}')$ denotes a two-particle interaction potential.
The perturbative evaluation of (\ref{can27}) is based on expanding the exponential with respect to the interacting action (\ref{can29}) in a power series, which reads up to first order
\begin{equation}
 \hspace*{-2mm}   Z_{N}(\beta)  = \frac{1}{N!} \int d\mathbf{x}_{1} \cdots \int d\mathbf{x}_{N}  \sum_{\text{P}} \prod_{n=1}^{N} \int_{\mathbf{x}_{n}(0) = \mathbf{x}_{n}}^{\mathbf{x}_{n}(\hbar \beta) = \mathbf{x}_{\text{P}(n)}} \mathcal{D}\mathbf{x}_{n}(\tau)\; e^{-\mathcal{A}^{(0)}[\mathbf{x}_1, \ldots, \mathbf{x}_N]/\hbar} \left\{ 1 - \frac{1}{\hbar}\,\mathcal{A}^{(\text{int})}[\mathbf{x}_1, \ldots, \mathbf{x}_N]+ \ldots \right\}\, .
    \label{can31}
\end{equation}
Inserting (\ref{can29}) into (\ref{can31}) yields, together with (\ref{can1}),
\begin{eqnarray}
 &&    Z_{N}(\beta) = Z_{N}^{(0)}(\beta) - \frac{1}{2\hbar N!}\sum_{\text{P}}  \int d\mathbf{x}_{1} \cdots\int d\mathbf{x}_{N}\sum_{n' \neq m' = 1}^{N}
 \int_{0}^{\hbar \beta} d\tau \nonumber \\ &&\times  \prod_{n=1}^{N} \int_{\mathbf{x}_{n}(0) = \mathbf{x}_{n}}^{\mathbf{x}_{n}(\hbar \beta) = \mathbf{x}_{\text{P}(n)}}  \mathcal{D}\mathbf{x}_{n}(\tau)\,
 V^{(\text{int})}\Big(\mathbf{x}_{n'}(\tau) - \mathbf{x}_{m'}(\tau)\Big) 
  e^{-\mathcal{A}^{(0)}[\mathbf{x}_1, \ldots, \mathbf{x}_N]/\hbar}+ \ldots
   \, .
    \label{can32}
\end{eqnarray}
The remaining path integral deals with the summation over all possible paths of $N$ bosons between the initial time $\tau_{\rm i}=0$ and the final time $\tau_{\rm f}=\hbar \beta$, where a particle interaction takes place at some intermediate time $\tau$. Thus, we have to extend the perturbative path integral concept
for one particle \cite[Chap.~6]{feynman-path} to $N$ bosons. This means that we have to consider a free propagation both from the initial time $\tau_{\rm i}=0$ to the intermediate time $\tau$ and, subsequently, from the intermediate time $\tau$ to the final time 
$\tau_{\rm f}=\hbar \beta$ as well as perform ordinary integrals over all coordinates at which the interaction takes place. This converts (\ref{can32}) to
\begin{eqnarray}
    Z_{N}(\beta) &=& Z_{N}^{(0)}(\beta) -  \frac{1}{2\hbar N!} \sum_{\text{P}} \sum_{n \neq m = 1}^{N} \int_{0}^{\hbar \beta} d\tau \int d\mathbf{x}_{1} \cdots \int d\mathbf{x}_{N} \int d\mathbf{x}_{1}'' \cdots \int d\mathbf{x}''_{N} \nonumber \\ &&\times 
    (\mathbf{x}_{\text{P}(1)}, \ldots , \mathbf{x}_{\text{P}(N)} ;\hbar \beta|\mathbf{x}''_{1}, \ldots , \mathbf{x}''_{N} ; \tau)^{(0)} \, V^{(\text{int})}(\mathbf{x}''_{n} - \mathbf{x}''_{m}) \, (\mathbf{x}''_{1}, \ldots , \mathbf{x}''_{N}; \tau|\mathbf{x}_{1}, \ldots , \mathbf{x}_{N} ; 0)^{(0)}+ \ldots  \,.
    \label{can36}
\end{eqnarray}
Taking into account the factorization (\ref{can5})
of any many-particle propagator into a product of one-particle propagators yields
\begin{eqnarray}
    Z_{N}(\beta) &=& Z_{N}^{(0)}(\beta) -  \frac{1}{2\hbar N!} \sum_{\text{P}} \sum_{n \neq m = 1}^{N} \int_{0}^{\hbar \beta} d\tau \int d\mathbf{x}_{1} \cdots \int d\mathbf{x}_{N} \int d\mathbf{x}_{1}'' \cdots \int d\mathbf{x}''_{N}  \label{can37} \\ &&\times 
  (\mathbf{x}_{\text{P}(1)}; \hbar \beta|\mathbf{x}_{1}'', \tau)^{(0)}  \cdots (\mathbf{x}_{\text{P}(N)}; \hbar \beta|\mathbf{x}_{N}'', \tau)^{(0)}   \,V^{(\text{int})}  (\mathbf{x}''_{n} - \mathbf{x}''_{m})\; (\mathbf{x}_{1}''\, \tau|\mathbf{x}_{1}, 0)^{(0)} \cdots (\mathbf{x}_{N}''; \tau|\mathbf{x}_{N}, 0)^{(0)}
  + \ldots \, .
   \nonumber
\end{eqnarray}
Due to the indistinguishability of bosons, we also recognize in  Eq.~(\ref{can37}) that a symmetric permutation group summation is performed. But this time we have to take special care of how to treat the interaction term. Apart from the closed cycles, which appear in the non-interacting case,  
in addition interacting cycles occur in which the product of single-particle propagators connects the positions $\mathbf{x}_{n}''$, $\mathbf{x}_{m}''$, which are involved in the interaction $V^{(\text{int})}(\mathbf{x}_{n}'' - \mathbf{x}_{m}'')$. We recognize that there exist two different paths that connect two positions, where an interaction occurs. The direct path connects the interaction point $\mathbf{x}_m''$ with itself
\begin{equation}
     \alpha_{k}(\mathbf{x}_m'', \mathbf{x}_m''; \beta) = \int d\mathbf{x}_{1} \cdots \int d\mathbf{x}_{k}\;(\mathbf{x}_{m}''; \tau|\mathbf{x}_{1}; 0)^{(0)}(\mathbf{x}_{1}; \hbar \beta|\mathbf{x}_{2}; 0)^{(0)} \cdots (\mathbf{x}_{k-1}; \hbar \beta|\mathbf{x}_{k}; 0)^{(0)}(\mathbf{x}_{k}; \hbar \beta|\mathbf{x}_{m}''; \tau)^{(0)}\,.
    \label{can39}
\end{equation}
And the exchange path connects the two interactions points $\mathbf{x}_n''$ and
$\mathbf{x}_m''$ with $m \neq n$
\begin{equation}
   \gamma_{k}(\mathbf{x}_m'', \mathbf{x}_n''; \beta)= \int d\mathbf{x}_{1}\cdots  \int d\mathbf{x}_{k}\;(\mathbf{x}_{m}''; \tau|\mathbf{x}_{1}; 0)^{(0)}  (\mathbf{x}_{1}; \hbar \beta|\mathbf{x}_{2}; 0)^{(0)} \cdots (\mathbf{x}_{k-1}; \hbar \beta|\mathbf{x}_{k}; 0)^{(0)}(\mathbf{x}_{k}; \hbar \beta|\mathbf{x}_{n}''; \tau)^{(0)}\, .
     \label{can40}
\end{equation}
 Due to the spectral decomposition of the imaginary-time propagators (\ref{can5c}), both expressions (\ref{can39}) and (\ref{can40}) reduce to
\begin{eqnarray}
    \alpha_{k}(\mathbf{x}_m'', \mathbf{x}_m''; \beta) &=&  (\mathbf{x}_m'', k \hbar \beta|\mathbf{x}_{m}'',0)^{(0)}\, , \label{H3}\\
    \gamma_{k}(\mathbf{x}_m'', \mathbf{x}_n''; \beta) &= &(\mathbf{x}_m'', k \hbar \beta|\mathbf{x}_{n}'',0)^{(0)}\, ,
    \label{H3b}
\end{eqnarray}
respectively.
A pictorical representation of Eqs.~(\ref{H3}), (\ref{H3b}) for $k=3$ is shown in Fig.~\ref{Fig1}. 
With these considerations in mind, we can now further evaluate Eq.~(\ref{can37}).
Due to the interaction potential $V^{(\text{int})}(\mathbf{x}_{n}'' - \mathbf{x}_{m}'')$ we have to fix the two coordinates $\mathbf{x}_{n}''$, $\mathbf{x}_{m}''$ over which we have to integrate. From a combinatorial point of view
we can then have direct or exchange paths of lengths $l$ and $k-l$ involving the two coordinates $\mathbf{x}_{n}''$, $\mathbf{x}_{m}''$ as well as closed cycles for the remaining $N-k$ particles.
 The underlying combinatorics is exemplarily illustrated for $N=3$ particles in the Appendix \ref{derivation}. Generalizing the corresponding result (\ref{a9}) to the generic case of $N$ interacting particles
 by induction yields
\begin{eqnarray}
&&    Z_{N}(\beta) = Z_{N}^{(0)}(\beta) - \frac{\beta}{2} \sum_{k=2}^{N} \sum_{l=1}^{k-1} \int d\mathbf{x}_{m}''\int d\mathbf{x}_{n}''\,V^{(\text{int})}(\mathbf{x}_{m}'' - \mathbf{x}_{n}'')\, \Big[ \alpha_{l}(\mathbf{x}_m'', \mathbf{x}_m''; \beta) \alpha_{k-l}(\mathbf{x}_n'', \mathbf{x}_n''; \beta)\nonumber \\ 
    &&+ \gamma_{l}(\mathbf{x}_m'', \mathbf{x}_n''; \beta) \gamma_{k-l}(\mathbf{x}_n'', \mathbf{x}_m''; \beta) \Big] \, \frac{1}{(N-k)!} \int d\mathbf{x}_{1} \cdots \int d\mathbf{x}_{N-k} \sum_{\text{P}} \prod_{j=1}^{N-k} (\mathbf{x}_{P(j)}; \hbar \beta| \mathbf{x}_{j};0)^{(0)}+ \ldots\,.
    \label{can44}
\end{eqnarray}
After the bracket in the second line of Eq.~(\ref{can44}) we recognize  due to (\ref{can3}) and (\ref{can5}) that a non-interacting partition function for $N-k$ particles appears. 
Furthermore, the interacting cycles $\alpha$ and $\gamma$ lead, due to (\ref{H3}) and (\ref{H3b}), to energy contributions that correspond to the Hartree and Fock channel:
\begin{eqnarray}
     I_{l,k-l}^{({\rm H})}(\beta) &\equiv& \int d\mathbf{x}_{m}''\int d\mathbf{x}_{n}''\,V^{(\text{int})}(\mathbf{x}_{m}'' - \mathbf{x}_{n}'')\, (\mathbf{x}_m'', l \hbar \beta|\mathbf{x}_{m}'',0)^{(0)}\, (\mathbf{x}_n'', (k-l) \hbar \beta|\mathbf{x}_{n}'',0)^{(0)}\, ,
     \label{can45}\\
    I_{l,k-l}^{({\rm F})}(\beta) &\equiv& \int d\mathbf{x}_{m}''\int d\mathbf{x}_{n}''\,V^{(\text{int})}(\mathbf{x}_{m}'' - \mathbf{x}_{n}'')\,(\mathbf{x}_m'', l \hbar \beta|\mathbf{x}_{n}'',0)^{(0)}\, (\mathbf{x}_n'', (k-l) \hbar \beta|\mathbf{x}_{m}'',0)^{(0)} \, .
    \label{can46}
\end{eqnarray}
These energies involve two single-particle propagators, which are connected by the interacting potential and correspond to two different paths of lengths $l$ and $k-l$, respectively. The Hartree energy contains two direct paths, and the Fock energy consists of two exchange paths. Due to this interpretation, 
Hartree and Fock energies can be represented on a Feynman cylinder, as shown for $l=1$ and $k=3$ in Fig.~\ref{Fig2}. Another possibility of graphically illustrating Eqs.~(\ref{can45}) and (\ref{can46}) is depicted in Fig.~\ref{Fig3} in terms of Feynman diagrams. We recognize that they are identical with the Feynman diagrams appearing in first-order perturbative theory in the grand-canonical ensemble at high temperatures \cite{Glaum-many}, where the anomalous correlations turn out to be absent. 

But here in the canonical ensemble, we have to interpret the Feynman diagrams with different Feynman rules \cite{glaummaxplanck}. Each line corresponds to an imaginary-time propagator, where the attached number indicates the number of windings around the Feynman cylinder as illustrated
in Fig.~\ref{Fig2}.
With this, we arrive at the final formula for the interacting partition function $Z_{N}(\beta)$ up to the first order  \cite{jonata,glaummaxplanck}:
\begin{equation}
   Z_{N}^{(1)}(\beta) = Z_{N}^{(0)}(\beta) - \frac{\beta}{2} \sum_{k=2}^{N} \sum_{l=1}^{k-1} \left[I_{l,k-l}^{({\rm H})}(\beta) + I_{l,k-l}^{({\rm F})}(\beta) \right] Z_{N-k}^{(0)}(\beta) \,.
   \label{can47}
\end{equation}
Note that this first-order result can be quickly extended to higher orders. For instance, in second-order, we expect to have 5 topologically different diagrams, since the Feynman rules in the grand-canonical ensemble lead to 5 different diagrams \cite{Glaum-many}.

\begin{figure}

\[
I^{({\rm H})}_{\textcolor{red}{1}, \textcolor{green}{2}} = 
\quad (\textcolor{red}{1})\quad
\begin{tikzpicture}[baseline={(current bounding box.center)}]

\draw[thick] (0,0) circle (0.3);
\draw[thick, ->] (-0.3,0) arc (180:90:0.3);

\draw[thick] (1.2,0) circle (0.3);
\draw[thick, ->] (1.5,0) arc (0:90:0.3);

\draw[thick, decorate, decoration={snake, segment length=4pt, amplitude=1.5pt}]
(0.3,0) -- (0.9,0);

\fill (0.3,0) circle (1.5pt);
\fill (0.9,0) circle (1.5pt);

\node at (2.0,0.1) {$(\textcolor{green}{2})$};

\end{tikzpicture}
\quad , \quad
I^{({\rm F})}_{\textcolor{red}{1},\textcolor{green}{2}} = 
\quad
\begin{tikzpicture}[baseline={(current bounding box.center)}]

\draw[thick] (0,0) circle (0.4);

\draw[thick, ->] (-0.4,0) arc (180:90:0.4); 
\draw[thick, ->] (0.4,0) arc (0:-90:0.4);    

\draw[thick, decorate, decoration={snake, segment length=4pt, amplitude=1.5pt}]
(-0.4,0) -- (0.4,0.0);

\fill (-0.4,0.0) circle (1.5pt);
\fill (0.4,0.0) circle (1.5pt);

\node at (0.0,0.75) {$(\textcolor{red}{1})$};
\node at (0.0,-0.75) {$(\textcolor{green}{2})$};

\end{tikzpicture}
\quad .
\]

\caption{Feynman diagrams for canonical Hartree and Fock energies involving three particles. Each line corresponds to an imaginary-time propagator, where the attached number indicates the number of windings around the Feynman cylinder depicted in Fig.~\ref{Fig2}.}
\label{Fig3}
\end{figure}
\subsection{Density matrix}
\begin{figure}[h]
\begin{tikzpicture}

\draw (0,0) -- (0,5) -- (5,0) ;
\draw[dashed] (0,0) -- (5,0)  ;
\draw (1,0) -- (1,4);
\draw (2,0) -- (2,3);
\draw (3,0) -- (3,2);
\draw (4,0) -- (4,1);
\draw[dashed] (0,4) -- (1,4);
\draw[dashed] (0,3) -- (2,3);
\draw[dashed] (0,2) -- (3,2);
\draw[dashed] (0,1) -- (4,1);
\tkzDefPoint(0,0){00};
\tkzDefPoint(0,1){01};
\tkzDefPoint(0,2){02};
\tkzDefPoint(0,3){03};
\tkzDefPoint(0,4){04};
\tkzDefPoint(0,5){05};
\tkzDefPoint(1,0){10};
\tkzDefPoint(1,1){11};
\tkzDefPoint(1,2){12};
\tkzDefPoint(1,3){13};
\tkzDefPoint(1,4){14};
\tkzDefPoint(2,0){20};
\tkzDefPoint(2,1){21};
\tkzDefPoint(2,2){22};
\tkzDefPoint(2,3){23};
\tkzDefPoint(3,0){30};
\tkzDefPoint(3,1){31};
\tkzDefPoint(3,2){32};
\tkzDefPoint(4,0){40};
\tkzDefPoint(4,1){41};
\tkzDefPoint(5,0){50};
\tkzLabelPoint[left](00){{\small$1$}}
\tkzLabelPoint[left](04){{\small$N-2$}}
\tkzLabelPoint[left](05){{\small$N-1$}}
\tkzLabelPoint[below,yshift=-0.2mm](00){{\small$1$}}
\tkzLabelPoint[below,yshift=-0.2mm](10){{\small$2$}}
\tkzLabelPoint[below,yshift=-0.2mm](50){{\small$N-1$}}
\foreach \n in {00,01,02,03,04,05,10,11,12,13,14,
20,21,22,23,30,31,32,40,41,50}
  \node at (\n)[circle,fill,inner sep=1.5pt]{};
\tkzDefPoint(-0.1,2){k};
\tkzLabelPoint[left](k){$k$}
\tkzDefPoint(-0.1,3){vdots2};
\tkzLabelPoint[left](vdots2){\huge $\vdots$}
\tkzDefPoint(-0.1,1){vdots1};
\tkzLabelPoint[left](vdots1){\huge $\vdots$}
\tkzDefPoint(3.,-0.1){n};
\tkzLabelPoint[below, yshift=-0.0mm](n){$n$}
\tkzDefPoint(2.,-0.1){cdots1};
\tkzLabelPoint[below, yshift=0.5mm](cdots1){\huge $\cdots$}
\tkzDefPoint(4.,-0.1){cdots2};
\tkzLabelPoint[below, yshift=0.5mm](cdots2){\huge $\cdots$}
\end{tikzpicture}
\caption{Triangle for nested summation in $(n,k)$ plane. In Eq.~(\ref{eqint37})
the left-hand side corresponds to adding points following each vertical line and later summing all vertical lines, whereas the sum at right-hand side adds all points along each horizontal line and later sums over all the horizontal lines.}
\label{triangle}
\end{figure}
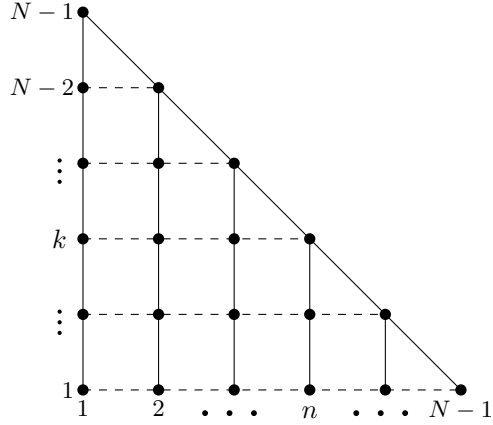

Applying the same procedure as in the previous section for calculating the $N$-particle partition function allows to determine also  the density matrix up to the first order in the interaction. The interaction version of Eq.~(\ref{can10c}) is given by
\begin{eqnarray}
\hspace*{-7mm}\rho_{N}(\mathbf{x}_1, \mathbf{x}'_1 ; \beta) = \frac{1}{N! Z_{N}(\beta)} \sum_{\text{P}} \int d\mathbf{x}_{2} \cdots \int d\mathbf{x}_{N}  \int_{\mathbf{x}_{1}(0) = \mathbf{x}_{1}'}^{\mathbf{x}_{1}(\hbar \beta) = \mathbf{x}_{\text{P}(1)}} \hspace*{-1mm}\mathcal{D}\mathbf{x}_{1}(\tau)
\prod_{n=2}^{N} 
\int_{\mathbf{x}_{n}(0) = \mathbf{x}_{n}}^{\mathbf{x}_{n}(\hbar \beta) = \mathbf{x}_{\text{P}(n)}} \hspace*{-1mm}\mathcal{D}\mathbf{x}_{n}(\tau)\;e^{-\mathcal{A}[\mathbf{x}_1, \ldots, \mathbf{x}_N]/\hbar}  \, .
    \label{can48}
\end{eqnarray}
Inserting the action (\ref{can28}) and
expanding the integrand with respect to the interacting contribution (\ref{can29}) up to first order, the zeroth order (\ref{can14}) is modified according to
\begin{eqnarray}
&&     \rho_{N}(\mathbf{x}_1, \mathbf{x}'_1 ; \beta) =  \frac{Z_{N}^{(0)}(\beta)}{Z_{N}(\beta)}
\rho_{N}^{(0)}(\mathbf{x}_1, \mathbf{x}'_1 ; \beta)
- \frac{1}{2 \hbar N! Z_{N}(\beta)} \sum_{\text{P}} \sum_{n \neq m = 1}^{N} \int d\mathbf{x}_2 \cdots \int \mathbf{x}_N  \int d\mathbf{x}''_1 \cdots \int d\mathbf{x}''_N \int_{0}^{\hbar \beta} d\tau
\label{can49} \\&&  \times 
(\mathbf{x}_{\text{P}(1)},\mathbf{x}_{\text{P}(2)}, \ldots , \mathbf{x}_{\text{P}(N)} ;\hbar \beta|\mathbf{x}''_{1},\mathbf{x}''_{2}, \ldots , \mathbf{x}''_{N} ; \tau)^{(0)}\, V^{(\text{int})}(\mathbf{x}''_n - \mathbf{x}_m'')\,(\mathbf{x}''_{1},\mathbf{x}''_{2}, \ldots , \mathbf{x}''_{N}; \tau|\mathbf{x}_{1}',\mathbf{x}_{2}, \ldots , \mathbf{x}_{N} ; 0)^{(0)}+\ldots \, ,
    \nonumber
\end{eqnarray}
where both $N$-particle propagators factorize via (\ref{can5}).
The interaction term in Eq.~(\ref{can49}) has a structure similar to the $N$-particle partition function $Z_{N}(\beta)$ in (\ref{can36}). But 
now one coordinate is excluded from the integration, and instead the initial and final positions 
$\mathbf{x}_1$ and $\mathbf{x}'_1$ are fixed. Combinatorially evaluating (\ref{can49}) we fix, as in Section \ref{part-weak}, the two coordinates $\mathbf{x}''_m$ and $\mathbf{x}''_n$, where the two-body interaction occurs.
Again we have either direct or exchange paths of lengths $l$ and $k-l$ involving the two coordinates $\mathbf{x}_{n}''$, $\mathbf{x}_{m}''$. But now we have for the remaining $N-k$ particles
open cycles of length $a$ and
and closed cycles of length $N-k-a$
similar to Section \ref{dens-zero}. This finally yields 
\begin{eqnarray}
  &&   \rho_{N}(\mathbf{x}_1, \mathbf{x}'_1 ; \beta) = \frac{Z_{N}^{(0)}(\beta)}{Z_{N}(\beta)}\rho_{N}^{(0)}(\mathbf{x}_1, \mathbf{x}'_1 ; \beta)
      - \frac{\beta}{2 Z_{N}(\beta)} \sum_{k=2}^{N} \sum_{l=1}^{k-1} \frac{1}{N-k} \int d\mathbf{x}_{n}''\;\int d\mathbf{x}_{m}''\;V^{(\text{int})}(\mathbf{x}''_n - \mathbf{x}_m'') \Big[ (\mathbf{x}_m''; l \hbar \beta| \mathbf{x}_m''; 0)^{(0)} \nonumber \\ &&\times \,(\mathbf{x}_n''; (k-l)\hbar \beta|\mathbf{x}_n''; 0)^{(0)} + (\mathbf{x}_m''; l \hbar \beta| \mathbf{x}_n''; 0)^{(0)}\,(\mathbf{x}_n'', (k-l) \hbar \beta| \mathbf{x}_m''; 0)^{(0)} \Big]\, \sum_{a=1}^{N-k}
      (\mathbf{x}_1, a \hbar \beta| \mathbf{x}_1',0)^{(0)} 
      Z_{N-k-a}^{(0)}(\beta)+ \ldots \, .
    \label{can50}
\end{eqnarray}
The integrals over $\mathbf{x}''_n$ and $\mathbf{x}''_m$ can be identified with the Hartree and Fock energies (\ref{can45}) and (\ref{can46}), so Eq.~(\ref{can50}) reduces to the final expression for the canonical one-particle density matrix up to first order in the interaction:
\begin{eqnarray}
 \rho_{N}(\mathbf{x}_1, \mathbf{x}'_1 ; \beta) &=& \frac{Z_{N}^{(0)}(\beta)}{Z_{N}(\beta)} \rho_{N}^{(0)}(\mathbf{x}_1, \mathbf{x}'_1 ; \beta)
     - \frac{\beta}{2 Z_{N}(\beta)} \sum_{k=2}^{N} \sum_{l=1}^{k-1} \frac{1}{N-k}  \Big[ I_{l,k-l}^{(\text{H})}(\beta) + I_{l,k-l}^{(\text{F})}(\beta)\Big]\nonumber \\ && \times \sum_{a=1}^{N-k}\;(\mathbf{x}_1, a \hbar \beta| \mathbf{x}_1',0)^{(0)} Z_{N-k-a}^{(0)}(\beta)+ \ldots  \, .
    \label{can51}
\end{eqnarray}
As in the non-interacting case (\ref{can15}) also the interacting one-particle density matrix is normalized, i.e., we have
\begin{equation}
   \int d\mathbf{x}_1\; \rho_{N}(\mathbf{x}_1, \mathbf{x}_1 ; \beta) = 1\, .
    \label{can52}
\end{equation}
With this and taking into account the recursive evaluation of the non-interacting partition function (\ref{can16})
for both $N$ and $N-k$ particles,
we obtain
\begin{equation}
Z_{N}(\beta)    = \frac{1}{N } \sum_{k=1}^{N} Z_{1}(k \beta)\,Z_{N-k}^{(0)}(\beta) - \frac{\beta}{2 } \sum_{k=2}^{N} \sum_{l=1}^{k-1}  \Big[ I_{l,k-l}^{(\text{H})}(\beta) + I_{l,k-l}^{(\text{F})}(\beta)\Big] Z_{N-k}^{(0)}(\beta)+ \ldots \, .
    \label{can53}
\end{equation}
The ideal partition function $Z_{N-n}^{(0)}(\beta)$ appearing in the first place on the right-hand side is now expressed  by $Z_{N-n}(\beta)$ due to Eq.~(\ref{can47}), so Eq.~(\ref{can53}) is rewritten as
\begin{eqnarray}
 \label{can55}    Z_{N}(\beta)&=& \frac{1}{N} \sum_{n=1}^{N} Z_{1}(n\beta)\,Z_{N-n}(\beta) + \frac{\beta}{2N} \sum_{n=1}^{N-1} Z_{1}(n\beta) \sum_{k=1}^{N-n} \sum_{l=1}^{k-1} \Big[ I_{l,k-l}^{(\text{H})}(\beta) + I_{l,k-l}^{(\text{F})}(\beta)\Big] Z_{N-n-k}^{(0)}(\beta)  \\
    &&- \frac{\beta}{2}\sum_{k=2}^{N} \sum_{l=1}^{k-1} \Big[ I_{l,k-l}^{(\text{H})}(\beta) + I_{l,k-l}^{(\text{F})}(\beta)\Big]  \frac{1}{N-k} \sum_{a=1}^{N-k} Z_{1}(a \beta)\,Z_{N-k-a}^{(0)}(\beta)+ \ldots \, .
  \nonumber  
\end{eqnarray}
Here we have used the freedom to substitute in the triple sum of the first line of Eq.~(\ref{can55}) the upper boundary $N$ of the $n$-sum by $N-1$ and the lower boundary $2$ of the $k$-sum by $1$, respectively. Furthermore, we observe that in the first line a nested summation with respect to the indices
$n$ and $k$, which can be interchanged. To this end, we identify all terms involved within the  $(n,k)$ plane, so the nested summation corresponds to all points in the triangle of Fig.~\ref{triangle}. Interchanging the order of the summations $n$ and $k$ gives then
\begin{equation}
    \sum_{n=1}^{N-1}\sum_{k=1}^{N-n} Z_{1}(n\beta)Z_{N-k-n}^{(0)}(\beta) = \sum_{k=1}^{N-1}\sum_{n=1}^{N-k} Z_{1}(n\beta)Z_{N-k-n}^{(0)}(\beta)
    \label{eqint37}
\end{equation}
Applying 
the recursion relation
(\ref{can16}) then yields 
\begin{equation}
     Z_{N}(\beta) = \frac{1}{N} \sum_{k=1}^{N} \left[Z_{1}(k\beta) - \frac{k \beta}{2} \sum_{l=1}^{k-1} \Big[ I_{l,k-l}^{(\text{H})}(\beta) + I_{l,k-l}^{(\text{F})}(\beta)\Big] \right] Z_{N-k}(\beta) + \ldots \,,
    \label{can57}
\end{equation}
which is exact up to the first order in the two-particle interaction.
This is the recursion formula already obtained in Refs.~\cite{glaummaxplanck,jonata}. It has the obvious problem for low temperatures $\beta \rightarrow \infty$, where $Z_{N}(\beta)$ becomes negative. In order to solve this problem, we perform a resummation as outlined in the following. 
\subsection{Ground-state Occupancy}\label{3C}
The last Subsection showed how the original recursion relation for the $N$-particle interacting partition function (\ref{can47}) can be resummed according to (\ref{can57}) within first-order perturbation theory. Now we work out a subsequent resummation, which not only improves thermodynamics in the canonical ensemble but also allows to straight-forwardly determine how the two-particle interaction affects the ground-state occupancy.

To this end, we insert the spectral representation of the single-particle propagator (\ref{can5c}) into the Hartree and Fock integrals (\ref{can45}) and (\ref{can46}), yielding for their sum
\begin{equation}
    I_{l,k-l}^{(\text{H})}(\beta) + I_{l,k-l}^{(\text{F})}(\beta) = \sum_{\mathbf{n}} \sum_{\mathbf{m}}V_{\mathbf{n}, \mathbf{m}}\, e^{-l \beta \varepsilon_{\mathbf{m}}-\beta(k-l)\varepsilon_{\mathbf{n}}}    \, ,\label{can62}
\end{equation}
where we introduced, for brevity, the matrix elements
\begin{equation}
    V_{\mathbf{n}, \mathbf{m}} = \int d\mathbf{x} \int d\mathbf{x}'\;V^{(\text{int})}(\mathbf{x} - \mathbf{x}') \Big[  |\psi_{\mathbf{n}}(\mathbf{x})|^2\,|\psi_{\mathbf{m}}(\mathbf{x}')|^2 + \psi_{\mathbf{n}}^{*}(\mathbf{x})\,\psi_{\mathbf{n}}(\mathbf{x}')\,\psi_{\mathbf{m}}^{*}(\mathbf{x})\,\psi_{\mathbf{m}}(\mathbf{x}') \Big] \, .
    \label{can63}
\end{equation}
Taking into account (\ref{can8b}) and (\ref{can62}) convert the recursion relation (\ref{can57})  up to the first order in the two-particle interaction to the final result
\begin{equation}
     Z_{N}^{\rm R}(\beta) = \frac{1}{N} \sum_{k=1}^{N} Z_1 (k,\beta) Z_{N-k}^{\rm R}(\beta) \, .
    \label{can64}
\end{equation}
Although this formally resembles the non-interacting recursion relation (\ref{can16}), it contains interaction contributions via the effects one-particle partition function  $Z_1(k \beta)$ with $k$ windings around the Feynman cylinder is substituted by 
\begin{equation}
    Z_1(k,\beta) = \sum_{\mathbf{n}} e^{-k\beta E_{\mathbf{n}}(k)}\,.
    \label{can65}
\end{equation}
Here the energy $E_{\mathbf{n}}(k)$ contains apart from the one-particle energy $\varepsilon_{\mathbf{n}}$ also the matrix elements (\ref{can63}) according to
\begin{equation}
    E_{\mathbf{n}}(k) = \varepsilon_{\mathbf{n}} + \frac{1}{2} \sum_{\mathbf{m}} \sum_{l=1}^{k-1}\,V_{\mathbf{n}, \mathbf{m}}\,e^{-l\beta(\varepsilon_{\mathbf{m}} - \varepsilon_{\mathbf{n}})} \,.
    \label{can66}
\end{equation}
Now we are interested in a certain state $\mathbf{q}$, so (\ref{can65}) can be rewritten as 
\begin{equation}
Z_1 (k,\beta) = \Tilde{\gamma}_{\mathbf{q}} (k,\beta) + \xi_{\mathbf{q}} 
(k, \beta)\, .
\label{can67}
\end{equation}
Here the contribution from the specific state $\mathbf{q}$ reads
\begin{equation}
\Tilde{\gamma}_{\mathbf{q}} (k,\beta) = e^{-k\beta E_{\mathbf{q}}(k)}\, ,
    \label{can68}
\end{equation}
whereas the rest 
\begin{equation}
    \xi_{\mathbf{q}}(k,\beta) = \sum_{\mathbf{n} \neq \mathbf{q}} e^{-k\beta E_{\mathbf{n}}(k)}
    \label{can68b}
\end{equation}
excludes this state $\mathbf{q}$.
Inserting (\ref{can67}) into (\ref{can64}) yields
\begin{equation}
     Z_{N}^{\rm R}(\beta) = \frac{1}{N} \sum_{k=1}^{N} 
     \Big[
\Tilde{\gamma}_{\mathbf{q}} (k,\beta) + \xi_{\mathbf{q}} 
(k, \beta)
     \Big]
     Z_{N-k}^{\rm R}(\beta) \, ,
    \label{can68c}
\end{equation}
which allows us to determine the statistical properties of the specific state $\mathbf{q}$. Namely, as (\ref{can68})
obeys the property 
\begin{equation}
\Tilde{\gamma}_{\mathbf{q}} (k,\beta) = \left[ e^{-\beta E_{\mathbf{q}}(k)}\right]^k\, ,
    \label{can68d}
\end{equation}
we read off from (\ref{can68c}) that
\begin{equation}
    W_{\mathbf{q}}(k|N,\beta) = \Tilde{\gamma}_{\mathbf{q}} (k,\beta) 
    \,Z_{N-k}(\beta)
    \label{can69}
\end{equation}
represents the weight of finding {\it at least} $k$ particles in state $\mathbf{q}$. Thus, the probability of having {\it exactly} $k$ particles in state $\mathbf{q}$ is given by
\begin{equation}
    p_{\mathbf{q}}(k|N,\beta) = \frac{1}{Z_{N}^{\rm R}(\beta)}
\Big[  \Tilde{\gamma}_{\mathbf{q}} (k,\beta)
    \,Z_{N-k}^{\rm R}(\beta)
-    \Tilde{\gamma}_{\mathbf{q}} (k+1,\beta) 
    \,Z_{N-k-1}^{\rm R}(\beta)
\Big], .
    \label{can71}
\end{equation}
As a particularly interesting situation, let us specialize to the case $\mathbf{q} = \mathbf{0}$. Then (\ref{can71}) represents the recursion relation for the probability of finding $k$ of the $N$ interacting particles in the ground state.
\subsection{Moments, Cumulants, and Thermodynamics}
Based on the obtained probability distributions we can now derive useful formulas for its respective moments and cumulants. To this end, we focus on the case of the ground state, due to its special role for the macroscopic quantum phenomenon of Bose-Einstein condensation. Due to the mathematical similarity between the non-interacting and the interacting probability formulas  (\ref{can26}) and (\ref{can71}), we proceed along similar lines as in Section \ref{2C}
to relate the respective moments and cumulants with the renormalized partition function. For instance, the equation for the moments has a similar form of the non-interacting case (\ref{can27a}):
\begin{equation}
    \langle k^l \rangle_{\mathbf q}^{N,\beta}= \sum_{k=1}^N k^l\, p_{\mathbf{q}}(k|N,\beta)  \,,
    \label{can72}
\end{equation}
where now we have the interacting probability (\ref{can71}). So, if it is inserted into (\ref{can72}), the moments yield
\begin{equation}
 \langle k^m \rangle_{\mathbf{q}}^{N,\beta}
 =  \sum_{k=1}^{N} \Big[ k^m - (k-1)^m \Big]
 \Tilde{\gamma}_{\mathbf{q}} (k, \beta)\,\frac{Z_{N-k}^{\rm R}(\beta)}{Z_{N}^{\rm R}(\beta)} \, ,
    \label{can73}
\end{equation}
which is now valid up to first order with respect to the two-particle interaction. Correspondingly, we conclude for the cumulants
\begin{equation}
  \langle \kappa^m \rangle_{\mathbf{q}}^{N,\beta} 
    =  \langle k^m \rangle_{\mathbf{q}}^{N,\beta}
 - \sum_{j=1}^{m-1} \binom{m-1}{j} \langle \kappa^{j} \rangle_{{\mathbf q}}^{N,\beta} \,
 \langle k^{m-j} \rangle_{\mathbf{q}}^{N,\beta}\, .
    \label{can74}
\end{equation}
For practical purposes, we approximately characterize the underlying probability distribution
by the first four cumulants, where we have 
\begin{equation}
    \kappa_{1}(k) = \langle k \rangle 
    \label{eq1}\, ,
\end{equation}
\begin{equation}
    \kappa_{2}(k) = \langle k^{2} \rangle - \langle k \rangle^2\, ,
    \label{eq2}
\end{equation}
\begin{equation}
    \kappa_{3}(k) = \langle k^{3} \rangle - 3 \langle k \rangle \langle k^{2} \rangle + 2 \langle k \rangle^3 ,
    \label{eq3}
\end{equation}
\begin{equation}
    \kappa_{4}(k) =  \langle k^{4} \rangle - 4\langle k^{3} \rangle \langle k \rangle + 12\langle k^{2} \rangle \langle k \rangle^2 - 3\langle k^{2} \rangle^2 - 6\langle k \rangle^4\, .
    \label{eq4}
\end{equation}
Here the respective moments are calculated from Eq.~(\ref{can73}):
\begin{equation}
     \langle k \rangle = \frac{1}{Z_{N}^{\rm R}(\beta)} \sum_{k=1}^{N} \Tilde{\gamma}_{\mathbf q}^{k}(\beta)\,Z_{N-k}^{\rm R}(\beta) \,,
    \label{eq5}
\end{equation}
\begin{equation}
     \langle k^2 \rangle = \frac{1}{Z_{N}^{\rm R}(\beta)} \sum_{k=1}^{N} (2k-1)\,\Tilde{\gamma}_{\mathbf q}^{k}(\beta)\,Z_{N-k}^{\rm R}(\beta) \, ,
    \label{eq6}
\end{equation}
\begin{equation}
     \langle k^3 \rangle = \frac{1}{Z_{N}^{\rm R}(\beta)} \sum_{k=1}^{N} (3k^2-3k + 1)\,\Tilde{\gamma}_{\mathbf q}^{k}(\beta)\,Z_{N-k}^{\rm R}(\beta) \, ,
    \label{eq7}
\end{equation}
\begin{equation}
     \langle k^4 \rangle = \frac{1}{Z_{N}^{\rm R}(\beta)} \sum_{k=1}^{N} (4k^3 - 6k^2 + 4k -1)\,\Tilde{\gamma}_{\mathbf q}^{k}(\beta)\,Z_{N-k}^{\rm R}(\beta) \, .
    \label{eq8}
\end{equation}
All these cumulants for the ground-state occupancy are useful to determine the location of the phase transition between the normal Bose gas and the Bose-Einstein condensate.
In particular, the first and second cumulant are known as the condensate fraction and the ground-state fluctuation, respectively. And they are known to be experimentally accessible via time-of-flight expansions \cite{Vibel_2024, Arlt_2026}.

The above derived canonical recursion relation (\ref{can64}) for the partition function up to first order in the interaction strength does not only allow to calculate statistical quantities for a weakly interacting Bose gas. In addition, it becomes also possible to work out a thermodynamic analysis, which is based on the entropy and the heat capacity in the canonical ensemble.

To this end, we remind that the canonical ensemble considers a system which exchanges heat with a thermal reservoir such that the number of particles and temperature are fixed. As a consequence, the system endures energy fluctuations, that turn out to be proportional to the heat capacity. In order to determine the heat capacity, we have to take into account that the canonical ensemble description is connected to thermodynamics by the Helmholtz free energy \cite{huang}
\begin{equation}
    F_{N}(\beta) = -\frac{1}{\beta}\,\ln Z_{N}^{\rm R}(\beta)\, ,
    \label{thermo1}
\end{equation}
which obeys
\begin{equation}
    F_{N}(T) = E_{N}(T) - TS_{N}(T)\,.
    \label{thermo2}
\end{equation}
Here $E_{N}(T)$ denotes the internal energy and $S_{N}(T)$ stands for the entropy. Due to the first law of thermodynamics for reversible processes,
infinitesimal changes of the Helmholtz freee energy are given by
\begin{equation}
    dF_{N} = -p_{N}\,dV - S_{N}\,dT\, .
    \label{thermo3}
\end{equation}
This implies that the entropy follows according to
\begin{equation}
    S_{N}(T) = -\left. \frac{\partial F_{N}(T)}{\partial T}\right|_{V} \,. 
    \label{thermo4}
\end{equation}
The property of the system to store heat is quantified by the heat capacity
\begin{equation}
    C_{N}(T) = T \left. \frac{\partial S_{N}(T)}{\partial T} \right|_{V}  ~.
    \label{thermo5}
\end{equation}
Taking into account (\ref{thermo1}), (\ref{thermo4}), and (\ref{thermo5}) the latter response coefficient follows directly from the canonical partition function by
\begin{equation}
    C_{N}(T) =  k_{\rm B}\,T  \left. \frac{\partial^2} { \partial T^2} \Big[T \ln Z_{N}^{\rm R}(T) \Big]\right|_V  ~.
    \label{thermo6}
\end{equation}

\section{Results for dilute Bose gas in box trap}
As an application of canonical perturbation theory, we study a weakly interacting dilute Bose gas in a three-dimensional box trap. To compute the renormalized $N$-particle partition function $Z_{N}^{\rm R}(\beta)$ according to Eq.~(\ref{can64}), it is necessary to determine the effective single-particle partition function (\ref{can65}) as well as the Hartree-Fock integrals $I_{l,k-l}^{(\text{D})}(\beta)$ and $I_{l,k-l}^{(\text{F})}(\beta)$ contained in the matrix element (\ref{can63}). These equations depend on the non-interacting energy eigenvalues and -functions. However, one has to take into account that the latter crucially depend on the boundary conditions used to describe the box. Generically, the periodic boundary conditions are theoretically imposed in view of performing the thermodynamic limit, whereas the Dirichlet boundary conditions are the ones realized in an experiment. Therefore, we restrict ourselves to the case of Dirichlet boundary conditions.
In case of a dilute gas at low-enough temperatures, it is physically justified to approximate the two-particle interaction  by the contact potential  $V^{(\text{int})}(\mathbf{x}) = g\,\delta(\mathbf{x})$, where the interaction strength $g= 4 \pi \hbar^2 a_{\rm s}/M$
is determined by the s-wave scattering length $a_{\rm s}$ \cite{pitaevskii}. Thus, the Hartree and Fock contributions to the matrix elements (\ref{can63})  turn out to be equal, yielding
\begin{equation}
    V_{\mathbf{n}, \mathbf{m}} = 2g \int d\mathbf{x}\,|\psi_\mathbf{n}(\mathbf{x})|^2 |\psi_\mathbf{m}(\mathbf{x})|^2 \,. 
    \label{p0}
\end{equation}
Here we read off that the Hartree-Fock energy depends on the energy eigenfunctions.
%
%
In case of Dirichlet boundary conditions for a box trap the one-particle energy eigenfunctions read \cite{griffiths}
\begin{equation}
    \psi_{\mathbf{n}}(\mathbf{x}) = \left(\frac{2}{L}\right)^{3/2} \prod_{j=1}^{3}\,\sin \biggl(\frac{\pi n_{j} x_{j}}{L} \biggr)~,
    \label{box1}
\end{equation}
where $n_{j} \in \mathbb{N}$ for all $j = 1, 2, 3$. Their corresponding energy eigenvalues are given by
\begin{equation}
    \varepsilon_{\mathbf{n}}^{\rm D} = \varepsilon_{\rm{D}} \mathbf{n}^2 ~,
    \label{box2}
\end{equation}
with the energy scale
\begin{equation}
    \varepsilon_{\rm{D}} = \frac{\pi^2 \hbar^2}{2ML^2} ~.
    \label{box3}
\end{equation}
Calculating the matrix element (\ref{p0}) for the Dirichlet boundary conditions we obtain
\begin{equation}
    V_{\mathbf{n}, \mathbf{m}}^{\rm D} = 2g \prod_{j=1}^{3} \int_{0}^{L} dx_j\,|\psi_{n_j}(x_j)|^2 |\psi_{m_j}(x_j)|^2 ~,
    \label{box4}
\end{equation}
where $\psi_{n_j}(x_j)$ stands for the one-dimensional wave function
\begin{equation}
    \psi_{n_j}(x_j) = \left(\frac{2}{L} \right)^{1/2} \sin \left(\frac{\pi n_j x_j}{L}\right) ~.
    \label{box5}
\end{equation}
With this, the integral in (\ref{box4}) is given by \cite{gradshteyn2014}
\begin{equation}
    \int_{0}^{L} dx_j\,|\psi_{n_j}(x_j)|^2 |\psi_{m_j}(x_j)|^2 = \frac{1}{L} \left(1 + \frac{1}{2} \delta_{n_j,m_j} \right) ~.
    \label{box6}
\end{equation}
Inserting (\ref{box6}) into (\ref{box4}) leads to
\begin{equation}
    V_{\mathbf{n}, \mathbf{m}}^{\rm D} = \frac{2g}{L^3} \prod_{j=1}^{3}\left(1 + \frac{1}{2} \delta_{n_j,m_j} \right) ~.
    \label{box7}
\end{equation}
And with Eqs.~(\ref{box3}) and (\ref{box7}), the interacting dispersion (\ref{can66}) leads to 
\begin{equation}
    E_{\mathbf{n}}^{\rm{D}}(k) = \varepsilon_{\rm{D}} \mathbf{n}^2 + \frac{g}{L^3} \sum_{l=1}^{k-1} \prod_{j=1}^{3} \left(\sum_{m_j=1}^{\infty} e^{-l \beta \varepsilon_{\rm{D}}(m_j^2 - n_j^2)} + \frac{1}{2}\right) ~.
    \label{box8}
\end{equation}
Finally, the recursion relation for the renormalized partition function (\ref{can64}) is given by
\begin{equation}
    Z_{N}^{\rm R, \rm D}(\beta) = \frac{1}{N} \sum_{k=1}^{N} \left(\sum_\mathbf{n} e^{-k \beta E_{\mathbf{n}}^{\rm{D}}(k)} \right) Z_{N-k}^{\rm R ,\rm D}(\beta) ~.
    \label{box9}
\end{equation}

\subsection{Statistical and Thermodynamical Properties}
\begin{figure}
    \centering
    \includegraphics[scale=0.40]{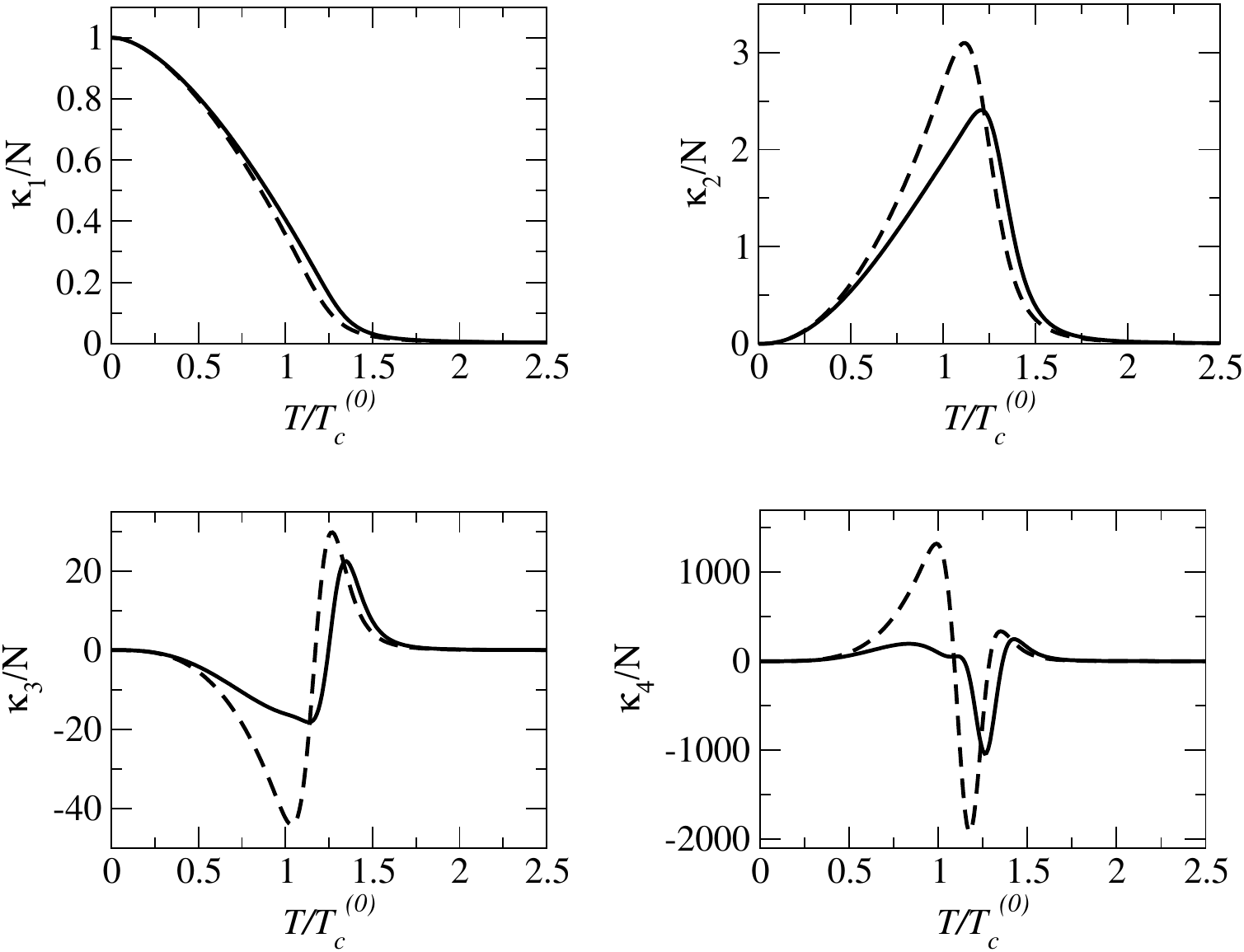}
    \caption{First four cumulants from Eqs.~(\ref{eq1})--(\ref{eq4}) for $500$ particles as function of dimensionless temperature $T/T_c^{(0)}$ with the grand-canonical critical temperature $T_c^{(0)}=(2 \pi \hbar^2/Mk_{\rm B})[N/L^3 \zeta(3/2)]^{2/3}$. Dashed lines are for the non-interacting case, i.e.~$an^{1/3}=0$, and  solid lines for the weakly interacting case $an^{1/3} = 0.1$ \cite{widera}.}
    \label{moments}
\end{figure}

\begin{figure}
    \centering
    \includegraphics[width=0.5\linewidth]{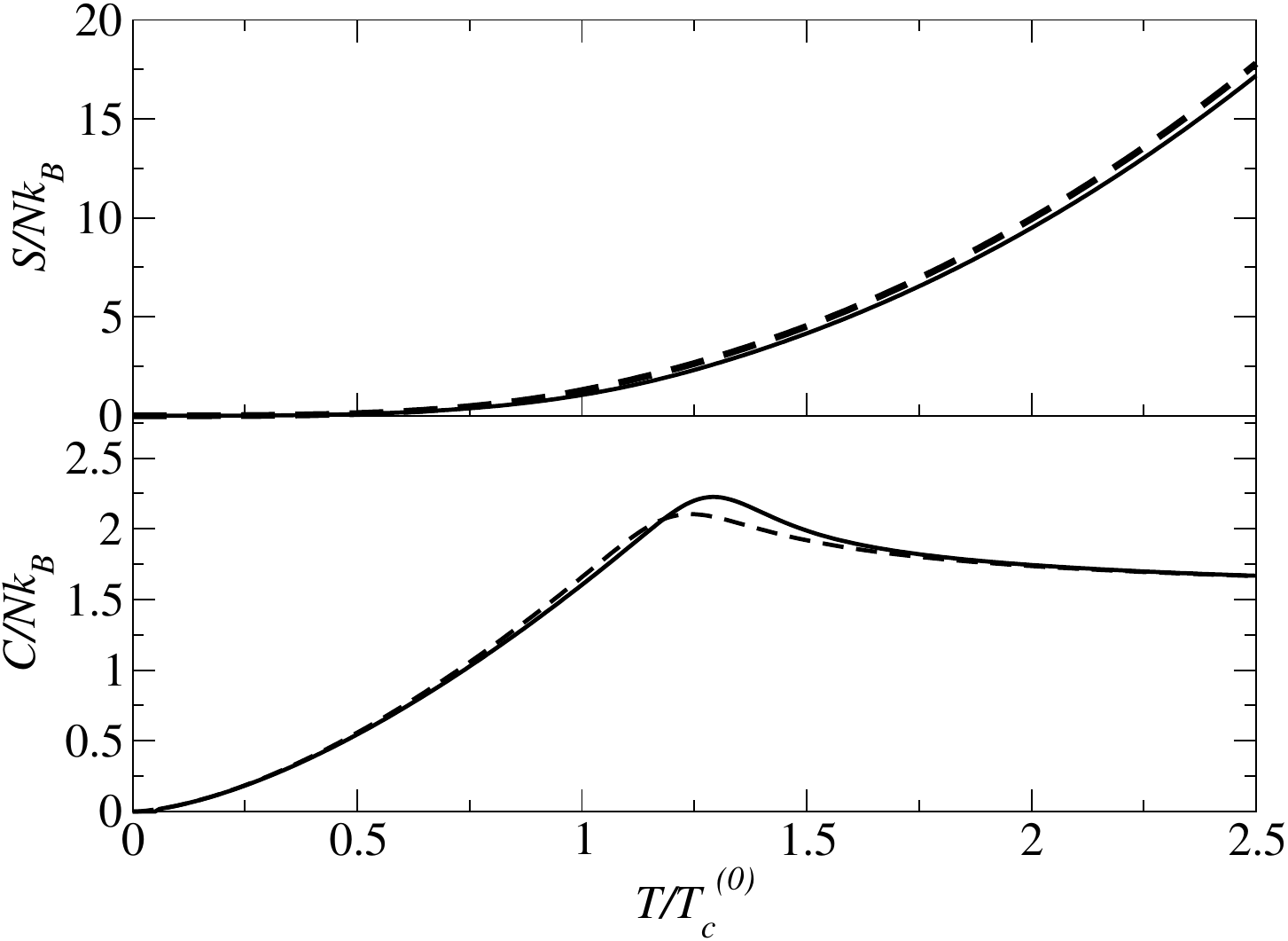}
    \caption{Entropy and heat capacity from Eqs.~(\ref{thermo4}), (\ref{thermo6}) for $500$ particles as function of dimensionless temperature $T/T_c^{(0)}$ with the grand-canonical critical temperature $T_c^{(0)}=(2 \pi \hbar^2/Mk_{\rm B})[N/L^3 \zeta(3/2)]^{2/3}$.
    Dashed lines are for the non-interacting case, i.e.~$an^{1/3}=0$, and  solid lines for the weakly interacting case $an^{1/3} = 0.1$ \cite{widera}.}
    \label{therm1}
\end{figure}

In Section \ref{3C} we worked out the moments and cumulants of the canonical probability distribution for occupying a certain state. They can be determined recursively for a weakly interacting system. As discussed previously, our focus is on the analysis of the ground-state statistics, thus Eqs. (\ref{eq5})--(\ref{eq8}) are considered for $\mathbf{q}  = 0$. In Section \ref{3C} we have also seen
that thermodynamic quantities can be calculated through the partition function. In the following, we compute the statistical and thermal properties of a dilute Bose gas for a box trap. The non-interacting results are recovered by putting the interacting potential equal to zero \cite{wilkens, glaum2007}.
The corresponding results for simulating 500 particles in a box are depicted in Figs.~\ref{moments} and \ref{therm1} for both an ideal and an interacting Bose gas with gas parameter $an^{1/3} = 0.1$, which corresponds to weak interactions.

Figure \ref{moments} shows the first four cumulants of the canonical probability distribution for particles being in the condensate,
where the dashed (solid) lines correspond to the ideal (interacting) results. Whereas the first cumulant represents the expectation value of the condensate, the second cumulant denoting the standard deviation is always positive. The third cumulant, which is also called skewness, can be either positive or negative. Negative (positive) skewness indicates larger (smaller) values than the expectation value and happens to occur for temperatures smaller (larger) than the critical temperature. Furthermore, the fourth cumulant, i.e.~the kurtosis, can be considered as a measure of the non-Gaussianity of the condensate. For a Gaussian random condensate density the kurtosis is zero; it is typically positive for distributions with heavy tails and a peak at zero, and negative for flatter densities with lighter tails. Distributions of positive or negative kurtosis, are thus called super-Gaussian or sub-Gaussian, respectively.
The ideal results for the first and second cumulants are the same as in Ref.~\cite{wilkens}, and the ideal third and fourth cumulants agree with  Ref.~\cite{scully2000}. With respect to thermodynamics, the ideal heat capacity is the same as in Ref.~\cite{glaum2007}. 

\begin{figure}
    \centering
    \includegraphics[width=0.5\linewidth]{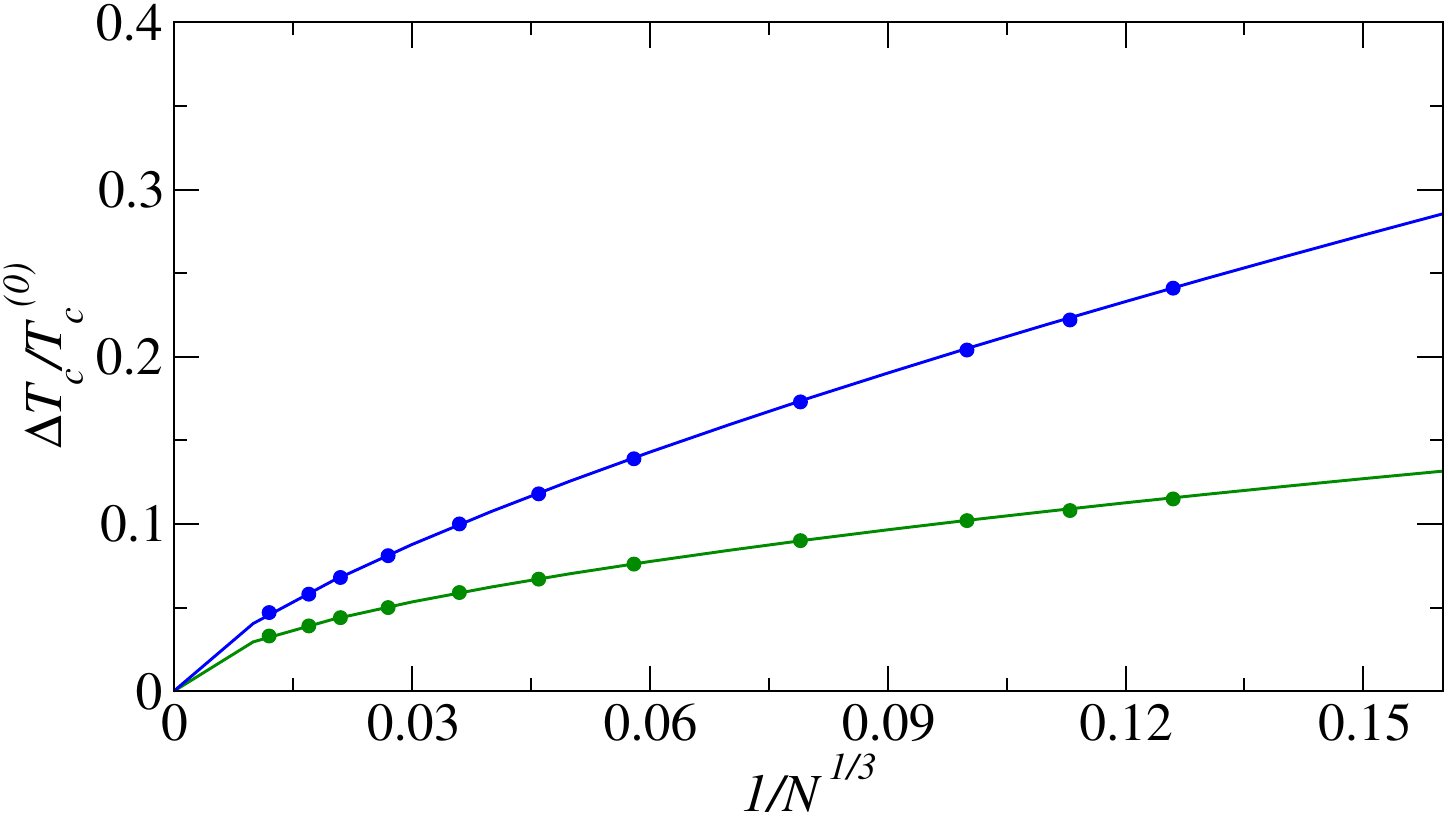}
    \caption{Critical temperature shift $\Delta T_c/T_{c}^{0}$ as function of $1/N^{1/3}$ calculated for the ideal Bose gas in Dirichlet case, where the blue circles correspond to the heat capacity (\ref{thermo6}) maximum points \cite{glaum2007}, and the green circles for the fluctuation (\ref{eq2}) maximum points. See that for small-$N$, the values have a big difference, but for large-$N$, they agree in the thermodynamic limit.}
    \label{tc_ideal}
\end{figure}

\begin{figure}
    \centering
    \includegraphics[width=0.5\linewidth]{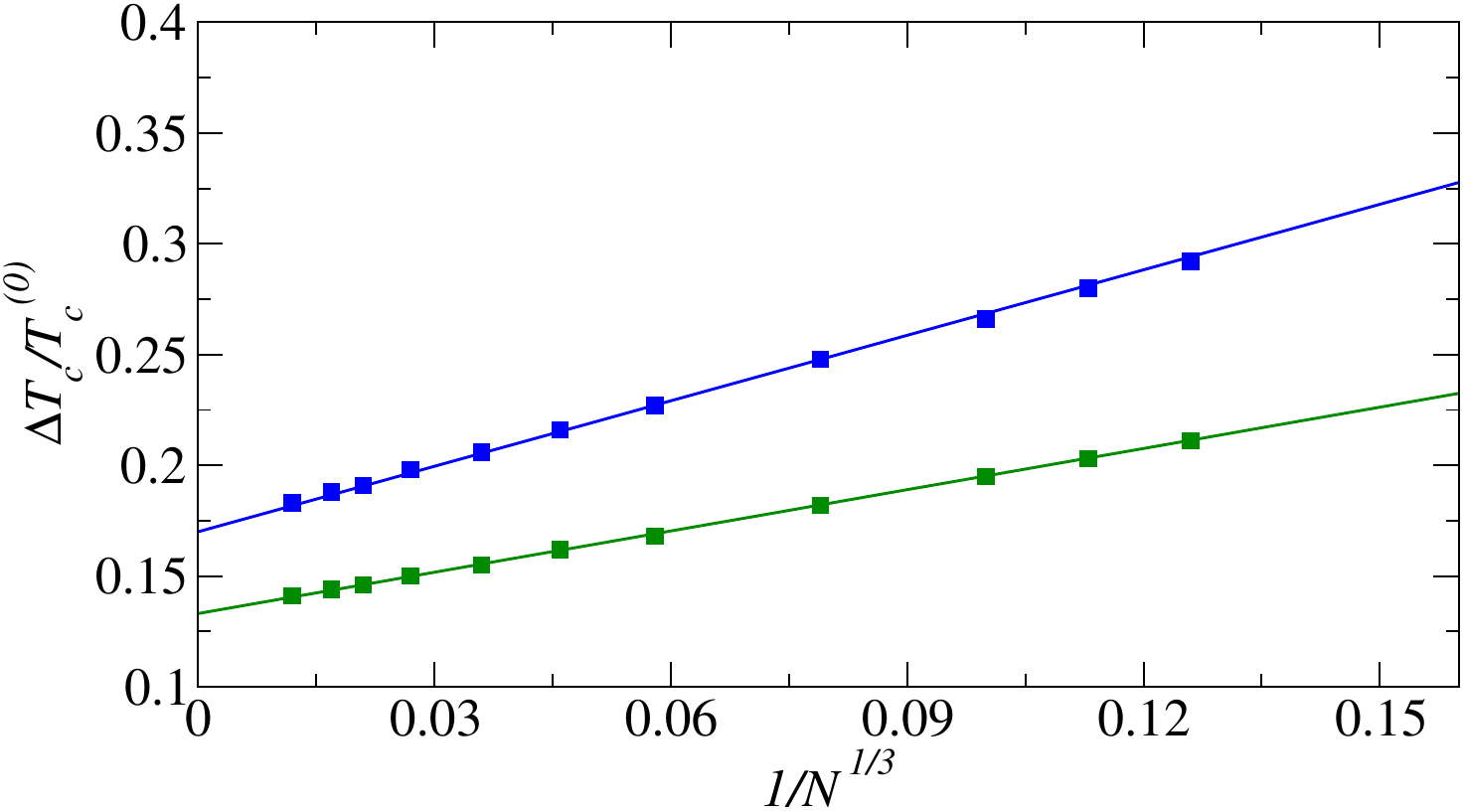}
    \caption{Critical temperature shift $\Delta T_c/T_{c}^{0}$ as function of $1/N^{1/3}$ calculated for the interacting Dirichlet case $a_s \rho^{1/3} = 0.1$, where the blue squares correspond to the heat capacity (\ref{thermo6}) maximum points, and the green squares for the fluctuation (\ref{eq2}) maximum points. See that for small-$N$, the values have a big difference, but for large-$N$, they have a fixed value, which by extrapolation, are $0.171$ - heat capacity, and $0.133$ - fluctuation.}
    \label{tc_int}
\end{figure}

From Figs.~\ref{moments} and \ref{therm1}, we can estimate the critical temperature by both the maximum points of ground-state fluctuation and the heat capacity plots. The results for the ideal case shown in Fig. \ref{tc_ideal} show that the critical temperatures obtained for each observable are different for smaller particles. But when $N$ increases, they converge to the same value, which can be identified with the thermodynamic limit. 
Including interactions with the gas parameter $a_s \rho^{1/3}$, we obtain the critical temperatures shown in Fig. \ref{tc_int}. Qualitatively, we recognize a similar behaviour as in the non-interacting case of Fig. \ref{tc_ideal}, but now 
the critical temperature determined from fluctuations turns out to be less than its heat capacity equivalent. Surprisingly, both cases converge to different results in the thermodynamic limit. For the heat capacity measure, we obtain $\Delta T_c/T_{c}^{0} = 0.171$, whereas for the fluctuation measure we get $\Delta T_c/T_{c}^{0} = 0.133$. The latter result coincides with the literature, where
both Quantum Monte Carlo simulations \cite{Moore_2001, Prokofiev_2001} and analytical calculations based on variational perturbation theory \cite{kleinert, kleinert_tc, Boris_tc} yield $\Delta T_c/T_c^{(0)}= 1.3 \,a_s n^{1/3}$. However, we note that our calculations were performed in the canonical ensemble, while the QMC simulations are implemented in the grand-canonical ensemble. Thus, we conclude that the fluctuation results obtained with the canonical approach converge in the thermodynamic limit to the grand-canonical description as in the case of an ideal Bose gas.  

Furthermore,
we read off  from Fig.~\ref{moments}a) that the interacting results calculated through canonical perturbation theory do not reproduce the quantum depletion at zero temperature. The reason for this is that the coupling constant expansion does not produce the non-perturbative result of the Bogoliubov theory \cite{Lee}, which is the correct result for low enough temperatures. However, the perturbation theory worked out in this article gives reliable interaction results for larger temperatures, in addition to taking into account the finite-size corrections. Furthermore, the behaviour of interacting quantities follows the predictions for larger temperatures \cite{kleinert, jonata}, computed previously in the grand-canonical ensemble or in a Bogoliubov canonical approach \cite{wang}. 
\section{Conclusions}
In this paper, we presented a canonical perturbation theory to calculate the thermodynamic and statistical properties of a weakly interacting Bose gas with contact interaction. The cycle decomposition method provides a diagrammatic expansion for canonical quantities which have the same Feynman diagrams as the grand canonical theory, but with different Feynman rules. The partition function as well as the moments and cumulants of the ground-state occupation turn out to
have respective recursive representations. However, as we obtained for low enough temperature the unphysical result of yielding a negative partition function, we worked out subsequently an appropriate resummation of the recursion formula. 
Finally, as an application of this theory, we studied the dilute Bose gas in a trap subject to Dirichlet boundary conditions with contact interaction. 

\begin{acknowledgments}
The authors are grateful to Konstantin Glaum, as he inspired this work through his PhD thesis \cite{glaumthesis}, which led to the articles \cite{glaum2007,Glaum-many,glaummaxplanck,dipolar}.
JSS and AG thank Conselho Nacional de Desenvolvimento Científico e Tecnológico (CNPq) grants nr. 141322/2024-1 and 306219/2022-0 for the financial support. AG also thanks the Brazilian agency Fundação de Amparo à Pesquisa do Estado de São Paulo (FAPESP)
grant nr. 2024/01533-7. AP acknowledges financial support from the Deutsche Forschungsgemeinschaft (DFG, German Research Foundation) through the Collaborative Research Center SFB/TR185 (Project No. 277625399). Furthermore, this work was supported by DAAD-CAPES PROBRAL, Brazil Grant Number 88887.627948/2021-00.
\end{acknowledgments}

\begin{appendix}
\section{Derivation of interacting partition function for three particles}\label{derivation}
\begin{figure}[t]
    \centering
    \includegraphics[scale=0.5]{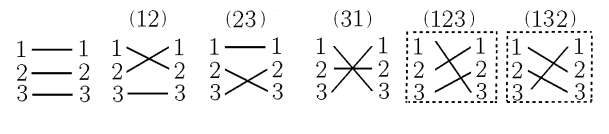}
    \caption{The possible permutations for three particles of different windings of the Feynman cylinder. For $N=3$, we have six possibilities of terms which satisfy the relation $C_1 + 2C_2 + 3C_3 = 3$, where $C_n$ is the number of terms with length $n$. }
    \label{permutation}
\end{figure}

Here we show exemplarity how the combinatorial calculation between Eq.~(\ref{can37}) and (\ref{can44}) is performed for three particles. Specializing Eq.~(\ref{can37}) for $N=3$, we have to evaluate
\begin{eqnarray}
    Z_{3}(\beta) &=&Z_{3}^{(0)}(\beta) - \frac{1}{12 \hbar} \sum_{n \neq m = 1}^{3} \sum_{\text{P}} \int_{0}^{\hbar \beta} d\tau \int d\mathbf{x}_{1} \cdots \int d\mathbf{x}_{3} \int d\mathbf{x}_{1}'' \cdots \int d\mathbf{x}_{3}''\;V^{(\text{int})}(\mathbf{x}_{n}'' - \mathbf{x}_{m}'')
     \\ && \times (\mathbf{x}_{\text{P}(1)}; \hbar \beta|\mathbf{x}_{1}'', \tau)^{(0)}  (\mathbf{x}_{\text{P}(2)}; \hbar \beta|\mathbf{x}_{2}'', \tau)^{(0)}(\mathbf{x}_{\text{P}(3)}; \hbar \beta|\mathbf{x}_{3}'', \tau)^{(0)} (\mathbf{x}_{1}''; \tau|\mathbf{x}_{1}, 0)^{(0)} (\mathbf{x}''_{2}; \tau|\mathbf{x}_{2}, 0)^{(0)} (\mathbf{x}_{3}''; \tau|\mathbf{x}_{3}, 0)^{(0)} \, .
    \label{a1}\nonumber
\end{eqnarray}
The underlying permutation group for $N=3$ has $3!=6$ elements, which are represented graphically in Fig.~\ref{permutation}. Thus, Eq.~(\ref{a1}) reads explicitly

\begin{eqnarray}
Z_3(\beta)& =& Z_3^{(0)}(\beta) - \frac{1}{12 \hbar} \sum_{n \neq m = 1}^{3}  \int_{0}^{\hbar \beta} d\tau \int d\mathbf{x}_{1}\int d\mathbf{x}_{2}\int d\mathbf{x}_{3} \int d\mathbf{x}_{1}''\int d\mathbf{x}_{2}''\int d\mathbf{x}_{3}''\label{a2} \\
&& \nonumber \times V^{(\text{int})}(\mathbf{x}_{n}'' - \mathbf{x}_{m}'') (\mathbf{x}_{1}''; \tau|\mathbf{x}_{1}, 0)^{(0)} (\mathbf{x}''_{2}; \tau|\mathbf{x}_{2}, 0)^{(0)} (\mathbf{x}_{3}''; \tau|\mathbf{x}_{3}, 0)^{(0)} \nonumber \\ 
&&\times \Big[(\mathbf{x}_{1}; \hbar \beta|\mathbf{x}_{1}'', \tau)^{(0)} (\mathbf{x}_{2}; \hbar \beta|\mathbf{x}_{2}'', \tau)^{(0)}(\mathbf{x}_{3}; \hbar \beta|\mathbf{x}_{3}'', \tau)^{(0)} + (\mathbf{x}_{1}; \hbar \beta|\mathbf{x}_{1}'', \tau)^{(0)} (\mathbf{x}_{3}; \hbar \beta|\mathbf{x}_{2}'', \tau)^{(0)}(\mathbf{x}_{2}; \hbar \beta|\mathbf{x}_{3}'', \tau)^{(0)} \nonumber\\ 
&& + (\mathbf{x}_{3}; \hbar \beta|\mathbf{x}_{1}'', \tau)^{(0)} (\mathbf{x}_{2}; \hbar \beta|\mathbf{x}_{2}'', \tau)^{(0)}(\mathbf{x}_{1}; \hbar \beta|\mathbf{x}_{3}'', \tau)^{(0)} + (\mathbf{x}_{2}; \hbar \beta|\mathbf{x}_{1}'', \tau)^{(0)} (\mathbf{x}_{1}; \hbar \beta|\mathbf{x}_{2}'', \tau)^{(0)}(\mathbf{x}_{3}; \hbar \beta|\mathbf{x}_{3}'', \tau)^{(0)} 
\nonumber  \\ &&+ (\mathbf{x}_{3}; \hbar \beta|\mathbf{x}_{1}'', \tau)^{(0)} (\mathbf{x}_{1}; \hbar \beta|\mathbf{x}_{2}'', \tau)^{(0)}(\mathbf{x}_{2}; \hbar \beta|\mathbf{x}_{3}'', \tau)^{(0)} + (\mathbf{x}_{2}; \hbar \beta|\mathbf{x}_{1}'', \tau)^{(0)} (\mathbf{x}_{3}; \hbar \beta|\mathbf{x}_{2}'', \tau)^{(0)}(\mathbf{x}_{1}; \hbar \beta|\mathbf{x}_{3}'', \tau)^{(0)}\Big] \,,
    \nonumber
\end{eqnarray}
Solving the integrals over the coordinates $\mathbf{x}_1, \mathbf{x}_2$, and $\mathbf{x}_3$ with the group property (\ref{cangroup}) yields
\begin{eqnarray}
&& Z_3(\beta) = Z_3^{(0)}(\beta)    - \frac{1}{12 \hbar} \sum_{n \neq m = 1}^{3}  \int_{0}^{\hbar \beta} d\tau \int d\mathbf{x}_{1}''\,\int d\mathbf{x}_{2}''\,\int d\mathbf{x}_{3}''\;V^{(\text{int})}(\mathbf{x}_{n}'' - \mathbf{x}_{m}'')   \label{a3} \\
&&\left\{ (\mathbf{x}_{1}''; \tau + \hbar \beta|\mathbf{x}_{1}'', \tau)^{(0)} \Big[ (\mathbf{x}_{2}''; \tau + \hbar \beta|\mathbf{x}_{2}'', \tau)^{(0)}
(\mathbf{x}_{3}''; \tau + \hbar \beta|\mathbf{x}_{3}'', \tau)^{(0)}+ 
 (\mathbf{x}''_{3}; \tau+\hbar \beta|\mathbf{x}_{2}'', \tau)^{(0)}(\mathbf{x}''_{2}; \tau+\hbar \beta|\mathbf{x}_{3}'', \tau)^{(0)} \Big]\right. \nonumber\\ 
&& + (\mathbf{x}_{3}''; \tau+\hbar \beta|\mathbf{x}_{1}'', \tau)^{(0)} 
\Big[ (\mathbf{x}_{2}''; \tau+\hbar \beta|\mathbf{x}_{2}'', \tau)^{(0)}(\mathbf{x}_{1}''; \tau+\hbar \beta|\mathbf{x}_{3}'', \tau)^{(0)} 
+ (\mathbf{x}_{1}''; \tau+\hbar \beta|\mathbf{x}_{2}'', \tau)^{(0)}(\mathbf{x}_{2}''; \tau+\hbar \beta|\mathbf{x}_{3}'', \tau)^{(0)} \Big]
\nonumber \\
&&\left. + (\mathbf{x}''_{2}; \tau+\hbar \beta|\mathbf{x}_{1}'', \tau)^{(0)}\Big[ (\mathbf{x}_{1}''; \tau+\hbar \beta|\mathbf{x}_{2}'', \tau)^{(0)}(\mathbf{x}_{3}''; \tau+\hbar \beta|\mathbf{x}_{3}'', \tau)^{(0)}+ (\mathbf{x}_{3}''; \tau+\hbar \beta|\mathbf{x}_{2}'', \tau)^{(0)}(\mathbf{x}_{1}''; \tau + \hbar \beta|\mathbf{x}_{3}'', \tau)^{(0)}\Big]\right\} \, .\nonumber
\end{eqnarray}
This intermediate result further simplifies due to two reasons. On the one hand the translational invariance  with respect to the imaginary time (\ref{can8a}) the integrand no longer depends on $\tau$, so the $\tau$-integral can be directly performed. Furthermore, the summation of the two-body interaction amounts to
\begin{equation}
    \sum_{m \neq n = 1}^{3} V^{(\text{int})}(\mathbf{x}_{n}'' - \mathbf{x}_{m}'') = 2 \Big[V^{(\text{int})}(\mathbf{x}_{1}'' - \mathbf{x}_{2}'') + V^{(\text{int})}(\mathbf{x}_{1}'' - \mathbf{x}_{3}'') + V^{(\text{int})}(\mathbf{x}_{2}'' - \mathbf{x}_{3}'')\Big]\, ,
    \label{a4}
\end{equation}
where all three terms lead to the same result due to underlying permutation symmetry. And
solving the integrals with respect to $\mathbf{x}_{3}''$ in Eq.~(\ref{a3}), we obtain the Hartree and Fock energies (\ref{can45}) and (\ref{can46}), respectively.
With this we finally arrive at
\begin{equation}
     Z_3(\beta) = Z_3^{(0)}(\beta) - \frac{\beta}{2} \left\{  Z_{1}(\beta) \left[I_{1,1}^{({\rm H})}(\beta) + I_{1,1}^{({\rm F})}(\beta)\right] + \left[I_{1,2}^{({\rm H})}(\beta) + I_{1,2}^{({\rm F})}(\beta) \right] + \left[ I_{2,1}^{({\rm H})}(\beta) + I_{2,1}^{({\rm F})}(\beta) \right]\right\} \, ,
     \label{a9}
\end{equation}
which is equivalent to Eq.~(\ref{can47}) for $N=3$.

\end{appendix}

\bibliographystyle{apsrev4-2}
\bibliography{main}
\end{document}